\documentclass{aa}  

\usepackage{graphicx}
\usepackage{txfonts} 
\usepackage{subcaption}
\usepackage{makecell} 
\usepackage{placeins} 

\newcommand\tstrut{\rule{0pt}{2.4ex}}
\usepackage{textcomp} 
\usepackage{siunitx}  

\newcommand*\dif{\mathop{}\!\mathrm{d}}

\usepackage{natbib}
\usepackage{hyperref}

\begin{document}

   \title{Abell 1033: Radio halo and gently reenergized tail at 54\,MHz}

 \author{H. W. Edler
          \inst{1}
          \and
          F. de Gasperin\inst{1,2}
          \and
          G. Brunetti\inst{2}
          \and          
          A. Botteon\inst{2,3,4}
          \and
          V. Cuciti\inst{1}
          \and
          R. J. van Weeren\inst{3}
          \and
          R. Cassano\inst{2}
          \and
          T. W. Shimwell\inst{3,5}
          \and
          M. Brüggen\inst{1}
          \and 
          A. Drabent\inst{6}
          }

   \institute{Hamburger Sternwarte, University of Hamburg,
              Gojenbergsweg 112, D-21029, Hamburg, Germany\\
              \email{henrik.edler@hs.uni-hamburg.de}
              \and
              INAF - Istituto di Radioastronomia,
              via P. Gobetti 101, 40129, Bologna, Italy
              \and
              Leiden Observatory, Leiden University, PO Box 9513, 2300 RA Leiden, The Netherlands
              \and 
              Dipartimento di Fisica e Astronomia (DIFA), Università di Bologna, via Gobetti 93/2, 40129, Bologna, Italy
              \and 
              ASTRON, Netherlands Institute for Radio Astronomy, Oude Hoogeveensedijk 4, 7991 PD, Dwingeloo, The Netherlands 
              \and
              Th{\"u}ringer Landessternwarte, Sternwarte 5, D-07778 Tautenburg, Germany 
              }
   \date{Received April 8, 2022; accepted July 21, 2022}
  \abstract
   {Abell 1033 is a merging galaxy cluster of moderate mass ($M_{500} = 3.24 \times 10^{14} M_\odot$). It hosts a broad variety of diffuse radio sources that are linked to different astrophysical phenomena. The most peculiar phenomenon is an elongated feature with an ultra-steep spectrum that is the prototype of the category of gently reenergized tails (GReET). Furthermore, the cluster hosts sources that were previously classified as a radio phoenix and a radio halo.}
   {We aim to improve the understanding of the cosmic-ray acceleration mechanisms in galaxy clusters in a frequency and mass range that has been poorly explored so far. }
   {To investigate the ultra-steep synchrotron emission in the cluster, we performed a full direction-dependent calibration of a LOFAR observation centered at 54\,MHz. We analyzed this observation together with recalibrated data of the LOFAR Two-meter Sky Survey at 144\,MHz and an archival GMRT observation at 323\,MHz. We performed a spectral study of the radio galaxy tail that is connected to the GReET to test whether the current interpretation of the source agrees with observational evidence below 100\,MHz. Additionally, we employed a Markov chain Monte Carlo code to fit the halo surface brightness profile at different frequencies.}
   {
   We report an extreme spectral curvature for the GReET. The spectral index flattens from $\alpha_{144\mathrm{MHz}}^{323\mathrm{MHz}} \approx -4$ to $\alpha_{~54\mathrm{MHz}}^{144\mathrm{MHz}} \approx -2$. This indicates the presence of a cutoff in the electron energy spectrum. 
  At the cluster center, we detect the radio halo at 54, 144, and at lower significance at 323\,MHz. We categorize it as an ultra-steep spectrum radio halo with a low-frequency spectral index $\alpha = -1.65 \pm 0.17$. Additionally, with a radio power of $P_\mathrm{150\,MHz} = 1.22\pm 0.13 \times 10^{25}$\,W\,Hz$^{-1}$, it is found to be significantly above the correlations of radio power to cluster mass reported in the literature. Furthermore, the synchrotron spectrum of the halo is found to further steepen between 144 and 323\,MHz, in agreement with the presence of a break in the electron spectrum, which is a prediction of homogeneous reacceleration models.}
   {}
   \keywords{galaxies: clusters: individual: Abell 1033 -- radio continuum: general –- X-rays: galaxies: clusters
               }

\maketitle

\section{Introduction}

Galaxy clusters are the building blocks of the large-scale structures of the Universe. When observed at radio wavelengths, we can detect synchrotron emission that traces the nonthermal plasma in the cluster volume \citep[for a review, we refer to][]{vanWeeren2019}. Different phenomena are associated with this emission. There is consensus that diffuse cluster-scale sources such as radio halos develop in the intracluster medium (ICM) as a consequence of mergers between clusters \citep{Brunetti2014}. Radio halos are megaparsec-scale sources with a steep spectral index of $\alpha \lesssim -1.2$\footnote{We follow the $S\propto\nu^\alpha$ spectral index convention.} that are located in the central region of clusters, approximately following the thermal ICM brightness distribution as observed in the X-rays. The turbulent reacceleration model \citep{Brunetti2001,Petrosian2001,Brunetti2007,Miniati2015,Brunetti2016} is the favored scenario to explain the mechanism that generates radio halos. In this model, turbulence injected in the ICM by a merger event reaccelerates cosmic-ray electrons (CRe) from a population of at least mildly relativistic primary (e.g., accelerated by active galactic nuclei) or secondary (generated in proton-proton collisions) seed electrons. \citep{Brunetti2011,Pinzke2017}. This is supported by the observed connection between the dynamic state of galaxy clusters and radio halos \citep{Cassano2010,Cassano2013,Cuciti2015,Cuciti2021}. Another important expectation of this model is a cutoff in the electron spectrum that depends on the energetics of the merger event and translates into a gradual steepening in the observed synchrotron spectrum. This implies that at a frequency of $\sim 1$\,GHz, we would in general be able to detect radio halos that were generated during very energetic merger events, while at lower frequency, we might be sensitive to a population of radio halos that is characterized by very steep radio spectra ($\alpha<-1.5$), which are in general referred to as ultra-steep spectrum radio halos \citep[USSRH;][]{Cassano2006,Brunetti2008}. 
Such sources have been observed in only a small number of clusters so far \citep{Brunetti2008,Macario2011,vanWeeren2012,Wilber2018,Duchesne2021,diGennaro2021}. 

Other extended radio sources in galaxy clusters, such as gently reenergized tails \citep[GReETs;][]{deGasperin2017} and radio phoenixes \citep{Kale2011,Cohen2011,Duchesne2021}, are thought to trace ancient tails and lobes of radio galaxies that have been reaccelerated. 
They are smaller sources ($\sim100$\,kpc) and are directly connected to individual radio galaxies. The observational difference between radio phoenixes and GReETs lies in the morphology, the spectral properties, and in the presence of a shock in the ICM. These differences can be explained by a difference in the underlying reacceleration mechanism. Phoenixes are mostly irregular sources with a steep spectrum without clear spatial trends, and they are thought to be accelerated by the adiabatic compression of old radio lobes through a shock wave \citep{Ensslin2001,Ensslin2002}. In contrast, GReETs are elongated tails of radio galaxies with highly unusual properties. Starting from the host radio galaxy, spectral aging causes an increasing steepening of the observed synchrotron emission, but at some point, the tail shows an unexpected rebrightening that coincides with a constant or even flattening spectral index. This is thought to be explained by a gentle energetization mechanism that is probably connected to microturbulence in the tail that is induced by interactions with the ICM \citep{deGasperin2017,vanWeeren2021}. The exact nature of the mechanism remains unclear because only a small number of such sources or candidates have been reported in the literature so far, with very limited spectral information \citep{deGasperin2017,Cuciti2018,Wilber2018,Botteon2021b,Ignesti2022}.

Observations at low radio frequencies ($<1$\,GHz) are essential for studying ultra-steep sources such as USSRHs and GReETs because these sources remain undetectable at higher frequencies, which is due to their spectral properties. The LOw-Frequency Array \citep[LOFAR;][]{vanHaarlem2013} is the largest and most sensitive radio-interferometer operating at low frequencies. It is therefore perfectly suited for studying diffuse radio emission in clusters of galaxies.

Abell 1033 (PSZ2 G189.31+59.24, hereafter A1033) is a moderately massive ($M_{500} = 3.24^{+0.30}_{-0.32} \times 10^{14} M_\odot$) galaxy cluster at a redshift of  $z = 0.126$ \citep{Planck2016} that shows recent merger activity and hosts a steep spectrum source previously classified as a radio phoenix \citep{deGasperin2015}.
In addition, the cluster contains an elongated ultra-steep ($\alpha \approx -4.0$) source connected to a wide-angle tail radio galaxy (WAT).\ Gentle reacceleration has been suggested to explain the peculiar spectral properties \citep{deGasperin2017}. This source is the prototype of the GReET category, and it remains the most extreme of the reported cases of radio galaxy tails that show signs of reacceleration. Observations in the ultra-low frequency regime (< 100\,MHz) are necessary for a more detailed study of this source. Furthermore, a radio halo in the cluster has recently been reported in \citet{Botteon2022} based on 144\,MHz LOFAR observations.

We analyze a LOFAR low-band antenna (LBA) observation of A1033 and data of the LOFAR Two-Meter Sky-Survey \citep[LoTSS,]{Shimwell2017,Shimwell2019, Shimwell2022} to study the properties and spectral behavior of the peculiar radio sources of this cluster. Throughout this work, we assume a fiducial flat $\mathrm{{\Lambda}CDM}$ cosmology with $\Omega_\mathrm{m}=0.3$ and $H_0=70\,\mathrm{km\,s^{-1}\,Mpc^{-1}}$. All spatial distances are given with respect to a redshift of $z = 0.126$, at which one arcsecond corresponds to 2.53\,kpc.

\section{Observations and data reduction}

\begin{table}
\centering
\caption{Observations. }\label{tab:metadata}
\begin{tabular}{ l c c c c }\hline
Telescope & Obs. date & \thead{Time\\ $[$h$]$} & \thead[l]{Freq. \\ $[$MHz$]$} & \thead{Bandwidth \\ $[$MHz$]$} \tstrut \\ 
\hline \vspace{2mm}
LBA & 30 June 2018\,\, & 8 & 54 & 48 \tstrut \\
LOFAR HBA &  24 Nov. 2015$^\mathrm{1}$ & 8 & 144 & 48\\\vspace{2mm}
 & 24 Dec. 2018\,\, & 8 & 144 & 48 \\
 GMRT & 02 Nov. 2014$^\mathrm{1}$ & 5.5 & 323 & 32 \\ \hline
\end{tabular}
\label{tab:obs}
\tablebib{(1)~Originally published in \citet{deGasperin2017}.}
\end{table}

For the ultra-low frequency study of A1033, we reduced and analyzed a dedicated LOFAR LBA observation  (project LC10-008, PI: de Gasperin) between 30 and 78\,MHz. Furthermore, we used data of two 8\,h LOFAR HBA pointings from the LOFAR Two-metre Sky Survey \citep[]{Shimwell2017,Shimwell2019,Shimwell2022} centered at 144\,MHz, as well as archival data of a 5.5\,h GMRT observation at 323\,MHz and of a 64\,ks Chandra X-ray observation in the energy band 0.5-7 keV. 
The data reduction and analysis of the Chandra observation is described in \citet{deGasperin2015}, and for the reduction of the GMRT data we refer to \citet{deGasperin2017}. We summarize information about the radio observations in \autoref{tab:obs}

\subsection{LOFAR LBA data}

The LOFAR LBA observation was conducted in observation mode LBA\_OUTER, where the outer half of the station dipoles are used to minimize electromagnetic crosstalk. During the preprocessing, the data were flagged at a resolution of 1\,s and 3\,kHz to mitigate radio frequency interference. The data were subsequently averaged in time and frequency to \SI{4}{\second} and 49\,kHz. The multibeam capability of LOFAR LBA allowed it to continuously point one beam toward the calibrator source \emph{3C196} during the observation. This source is a well-behaved standard calibrator for the low band \citep{Heald2015}. The calibrator data were reduced with the LOFAR LBA calibrator pipeline described in \citet{deGasperin2019}. This pipeline is used to find solutions for station-based direction-independent effects as well as an initial estimate for the ionospheric phase errors. 
The calibrator solutions revealed a poor data quality for the station CS031LBA, and therefore, we excluded this station from further processing. The calibrator solutions were applied to the target field dataset. Additionally, we corrected the data for the primary beam toward the phase center.
Next, we performed direction-independent self-calibration of the target field to find solutions for the average ionospheric effects. We employed the self-calibration pipeline presented in \citet{deGasperin2020}. Starting from an initial model based on a collection of radio surveys
(\citet[][TGSS]{Intema2017},~\citet[][NVSS]{Condon1998},~\citet[][WENNS]{Rengelink1997},~\citet[][VLSSr]{Lane2014}), solutions for the direction-independent ionospheric total electron content (TEC), Faraday rotation, and second-order primary beam effect were derived in two rounds of self-calibration. 
The root mean square (rms) background noise of the self-calibrated image is $\SI{2.5}{mJy\,beam^{-1}}$ at a resolution of $35''$, the image quality is limited by significant direction-dependent ionospheric errors still present in the data. These errors were addressed in direction-dependent calibration, where we employed the calibration pipeline that is being developed for the LOFAR LBA Sky Survey \citep[LoLSS;][]{deGasperin2021} and based on the facet calibration strategy outlined in \citet{vanWeeren2016}. For the next processing steps, we reduced the data volume by averaging the data to a resolution of 8\,s in time.
We used the direction-independently calibrated image to isolate suitable direction-dependent calibrators. We employed the Python blob detector and source finder (PyBDSF; \citet{Mohan2015}) as source finder, and selected bright and compact sources by thresholding based on flux density and source area. Compact sources in a proximity of 6' at most were merged by a grouping algorithm. We selected all sources and groups of sources with a flux density above $S_\mathrm{min} > \SI{1.0}{Jy}\times\left({\nu}/{\SI{60}{\mega\hertz}}\right)^{-0.8}$ at $\nu=\SI{30}{\mega\hertz}$ or  $\nu=\SI{54}{\mega\hertz}$. The reason for this is that spectral variation, mainly of the primary beam, can make it hard to identify all bright sources at a single frequency. This yielded 29 calibrator directions for the first of two major iterations of our strategy. 
To prepare the data for further calibration, we subtracted the model obtained after direction-independent calibration to create a dataset that was empty up to model inaccuracies. Then, we looped through the calibrators ordered by their flux density at \SI{30}{\mega\hertz} and repeated the following steps. First, the model of the calibrator was readded to the empty data. Second, the data were phase shifted to the calibrator direction and subsequently averaged further to a resolution of 16\,s (if $S_{54} > 10\,\mathrm{Jy}$) or 32\,s (otherwise) in time and 0.39\,MHz in frequency. This averaging was possible because we only imaged the single calibrator and not the whole field of view (FoV), so that we were not limited by time or frequency smearing. Third, we corrected this averaged dataset for the primary beam in the new phase center. Then, we self-calibrated on this dataset: By imaging a small field around the calibrator, we obtained a model of the calibrator source(s). The data were smoothed in time and frequency with a baseline-dependent kernel size. Here, shorter baselines were smoothed with a larger kernel to increase the signal-to-noise ratio. We solved on the smoothed data for scalar phases against the model. The frequency interval of this solve step was \SI{0.39}{\mega\hertz}, the time interval varied from 64/128\,s in the first iteration down to 16/32\,s in the following minor cycles. These solutions were then applied to the data, and the calibrator direction was imaged again to obtain an improved model for the next self-calibration iteration. For sources brighter than 10\,Jy, we performed two additional diagonal amplitude calibration steps: one step that enforced constant solutions between stations to capture beam model errors, and a subsequent step solved to correct for remaining errors. If the rms noise of the image did not improve by more than 1\%\ during a self-calibration iteration and the noise level was lower than the initial value, we considered this direction to be converged. Otherwise, it was not further used as a calibrator direction. In this case, the solutions were discarded and the direction was divided between the neighboring calibrators. 
We resubtracted the converged calibrator model that was corrupted with the final solutions in that direction from the data to obtain a cleaner empty dataset. Artifacts in the image around the calibrator related to direction-dependent errors were now strongly reduced. This procedure was repeated for all 29 calibrator directions. 

\begin{figure*}[ht!]
\centering
\begin{subfigure}{0.49\textwidth}
    \centering
    \includegraphics[width=1.0\linewidth]{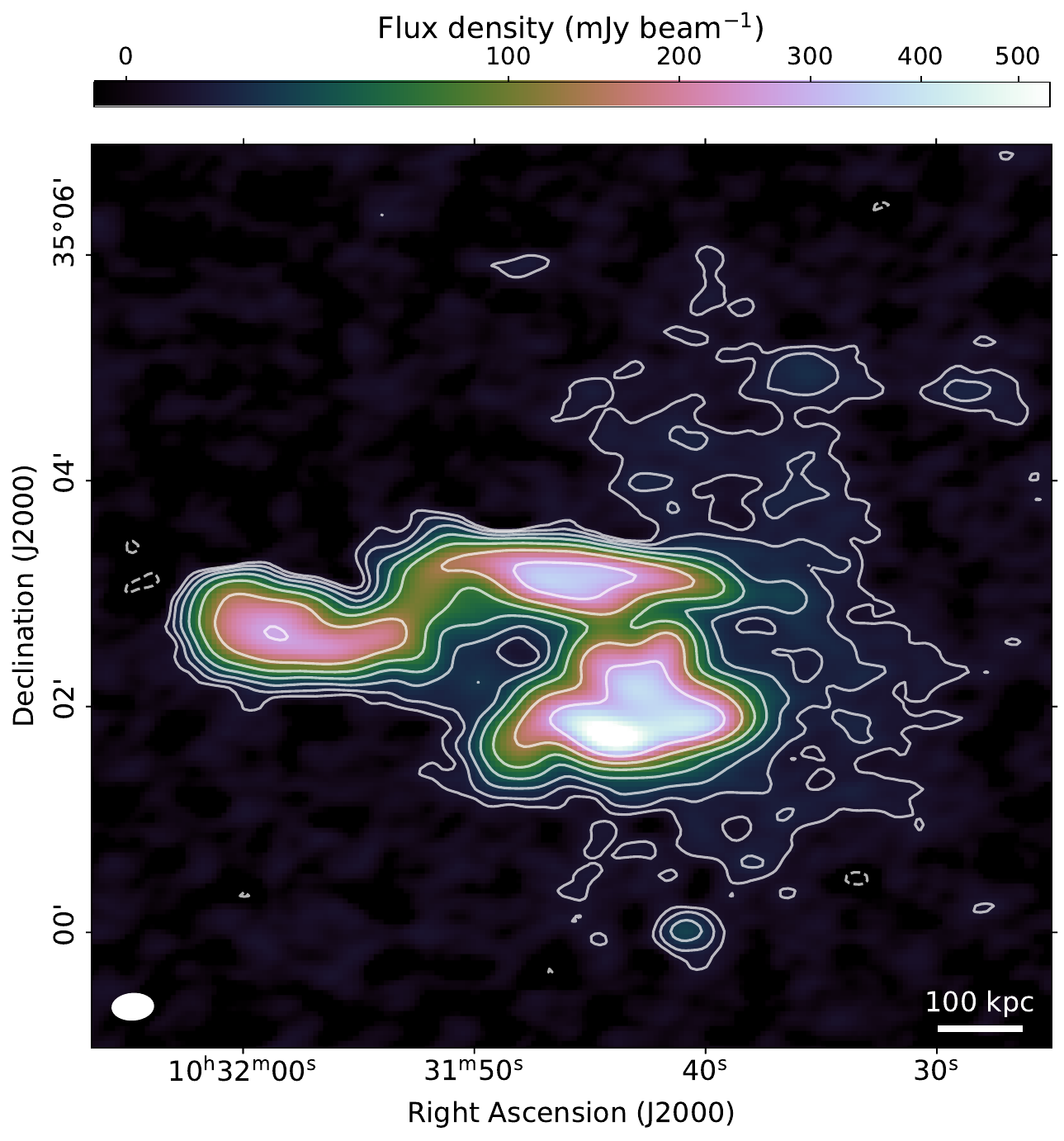}
    \caption{LBA 54\,MHZ, $21''\times14''$ resolution}
    \label{fig:imga}
\end{subfigure}
\begin{subfigure}{0.49\textwidth}
    \centering
    \includegraphics[width=1.0\linewidth]{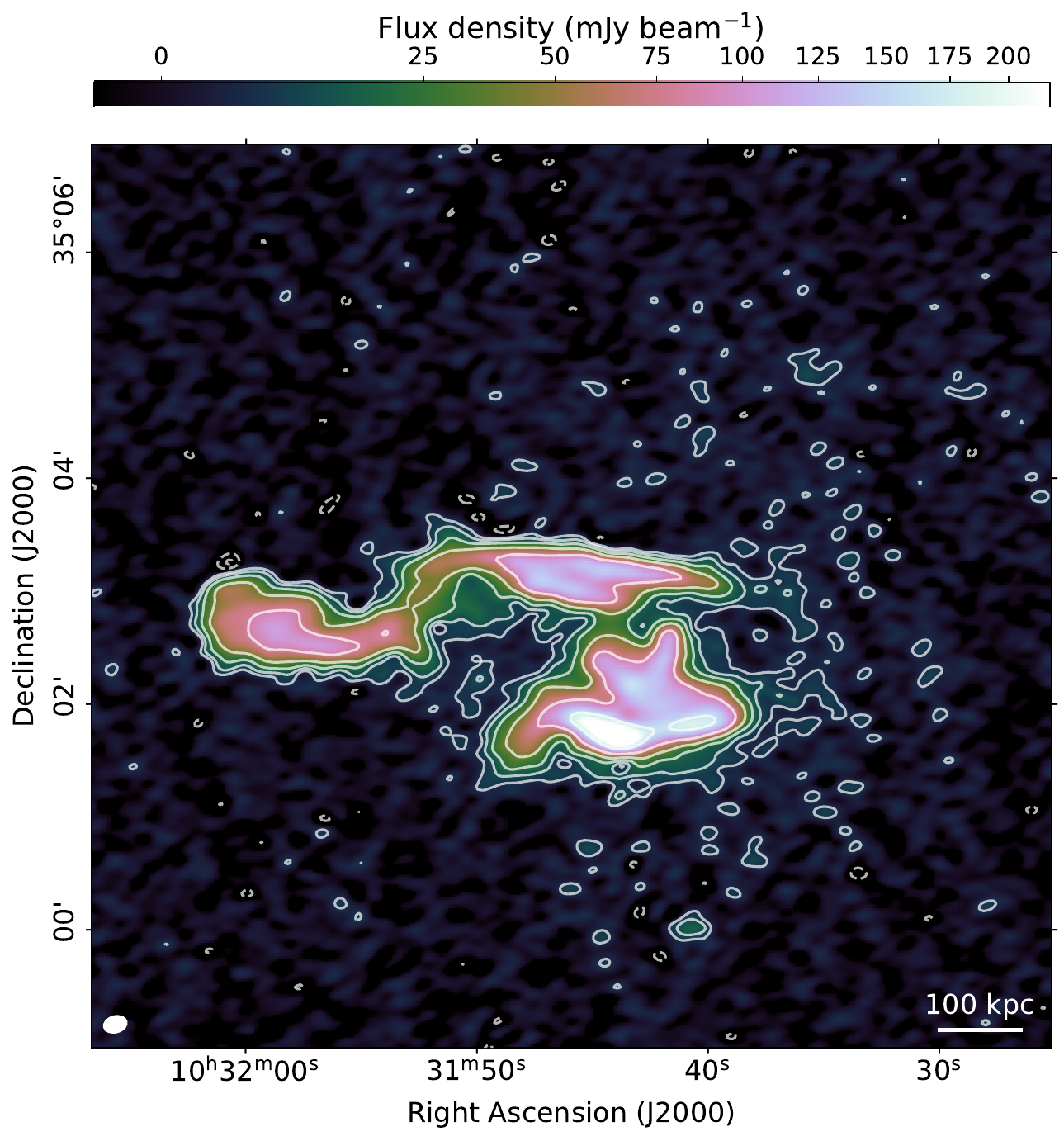}
    \caption{LBA 54\,MHZ, $12'' \times 9''$ resolution}
    \label{fig:imgb}
\end{subfigure}
\begin{subfigure}{0.49\textwidth}
    \centering
    \includegraphics[width=1.0\linewidth]{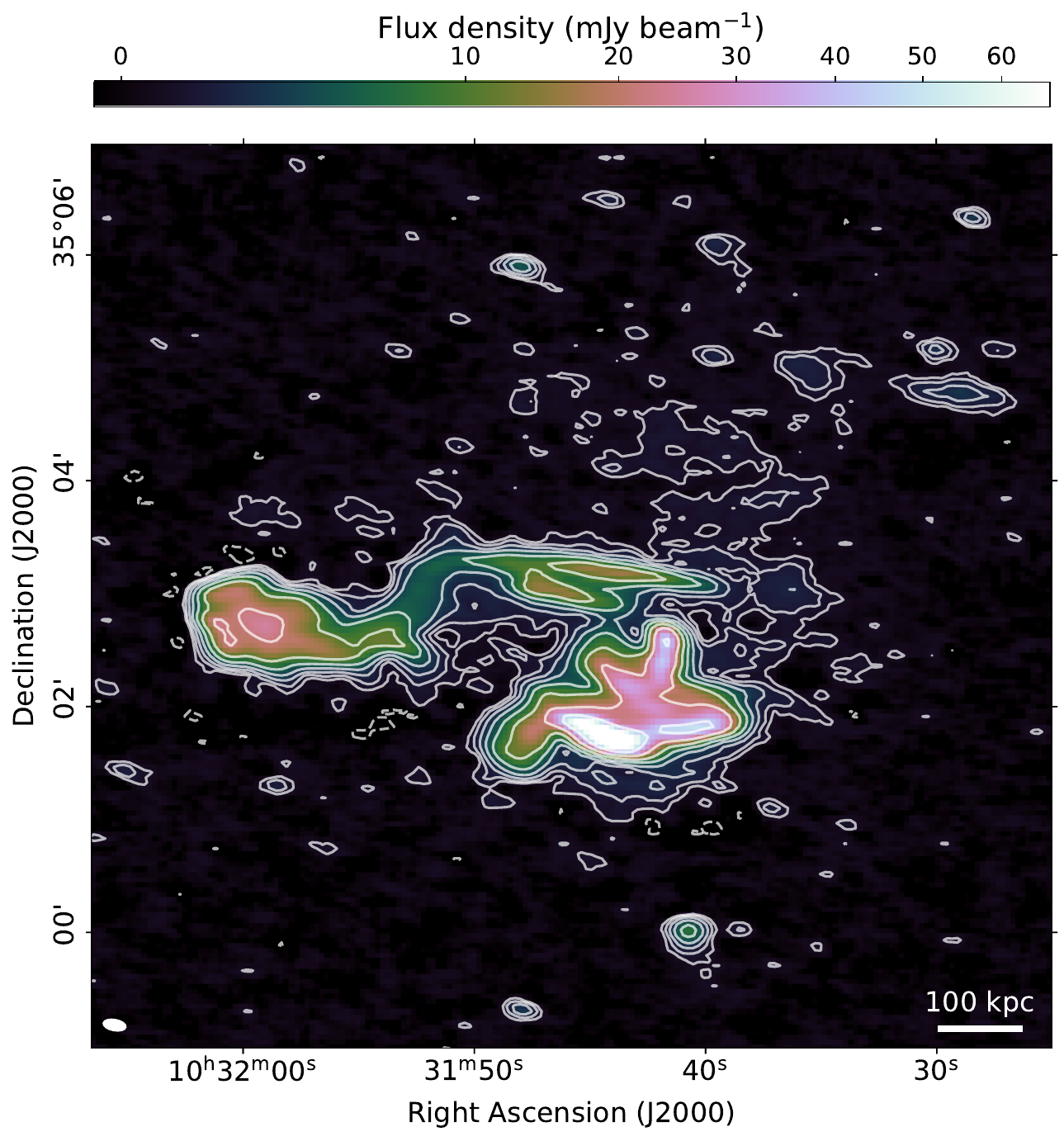}
    \caption{HBA 144\,MHZ, $12''\times6''$ resolution}
    \label{fig:imgc}
\end{subfigure}
\raggedright
\begin{subfigure}{0.49\textwidth}
    \centering
    \includegraphics[width=1.0\linewidth]{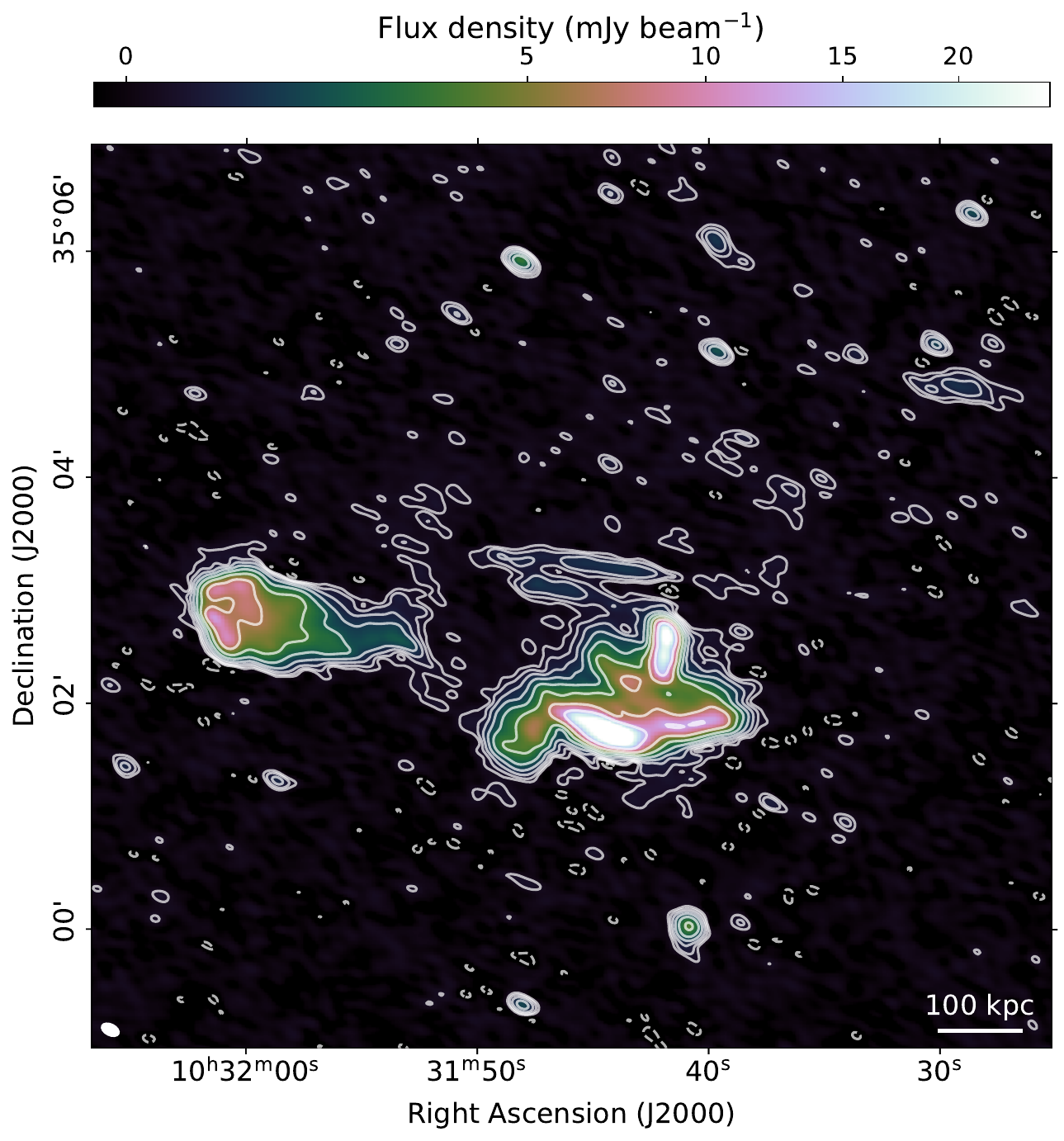}
    \caption{GMRT 330\,MHZ, $10''\times6''$ resolution}
    \label{fig:imgd}
\end{subfigure}
\caption{Four images (a) to (d) showing A1033 at different frequencies and resolutions. Contours start from $3\sigma$ and increase in powers of two. Dashed contours correspond to $-3\sigma$. The white ellipse in the bottom left corner of each image corresponds to the beam size as reported in \autoref{tab:images}. Image (d) was produced using archival GMRT data from \citet{deGasperin2017}.}
\label{fig:images}
\end{figure*}

We then started again with the averaged data after direction-independent calibration. From these data, we peeled\emph{} all calibrator sources outside of a 2.3$^\circ$ region around the pointing center, meaning that we subtracted the model after corrupting it with the corresponding direction-dependent solutions without readding it to the data at any point. This strongly reduced the impact of bright sources outside of the primary beam. These data were then imaged using \texttt{DDFacet} \citep{Tasse2018}. This imager simultaneously applied calibration solutions for the remaining 22 calibrator directions based on Voronoi-tesselated facets around the calibrators. The resulting image and model are the products of the first major iteration of our direction-dependent calibration pipeline. We performed a second major iteration using the improved model as input. This time, we searched for calibrators on the DDE-calibrated image and went 20\% lower (to 0.8\,Jy) in terms of minimum calibrator flux. We found 25 suitable directions for the second iteration, which were now all inside the primary beam main lobe. Again we looped over these sources ordered according to their brightness and repeated the individual self-calibration cycles as described above.
Since A1033 is a somewhat special source because it is both very bright (>10\,Jy at 54\,MHz) and extended (around 5'), our default algorithm had problems in correctly detecting all emission associated with A1033 as a calibrator direction. For this target, we therefore employd an LBA-specific implementation of the extraction-pipeline described in \citet{vanWeeren2021}. This pipeline subtracted all sources outside of a 0.2\textdegree\,radius around A1033 from the data using the model and calibration solutions from the direction-dependent pipeline. Then, it performed multiple rounds of self-calibration solving for TEC and a constant phase in each time interval. This \emph{\textup{tecandphase}}-constraint reduces the number of free parameters compared to the phase-solve used in direction-dependent calibration.
We imaged the final target-extracted dataset using a robust weighting of $-0.3$ and $-1.0$. The images have a background noise of 1.4 and 1.7\,\si{mJy\,beam^{-1}} at a resolution of $21''\times13''$ and $12''\times9''$. They are displayed in \autoref{fig:images}.

\begin{table}
    \centering
    \caption{Parameters of the images displayed in \autoref{fig:images}.  The Briggs value is the parameter for the baseline-dependent weighting of the visibility data during imaging.}
    \begin{tabular}{cccccc}
    Label & Telescope & \thead[c]{Freq.\\{[MHz]}} & Briggs & \thead{Resolution\\{[$''\times''$]}} & \thead[c]{Noise\\{[$\frac{\mathrm{mJy}}{\mathrm{beam}}$]}} \\\hline
    \hyperref[fig:imga]{Fig.\ref*{fig:imga}} & LBA & 54& $-0.3$ & $21\times14$ & 1.4 \\
    \hyperref[fig:imgb]{Fig.\ref*{fig:imgb}} & LBA & 54& $-1.0$ & $12\times9$ & 1.7 \\
    \hyperref[fig:imgc]{Fig.\ref*{fig:imgc}} & HBA & 144& $-0.3$ & $12\times6$ & 0.11 \\
    \hyperref[fig:imgd]{Fig.\ref*{fig:imgd}} & GMRT & 323& $-0.3$ & $10\times6$ & 0.034 \\
    \end{tabular}
    \label{tab:images}
\end{table}

\begin{figure}[ht!]
    \centering
    \includegraphics[width=1.0\linewidth]{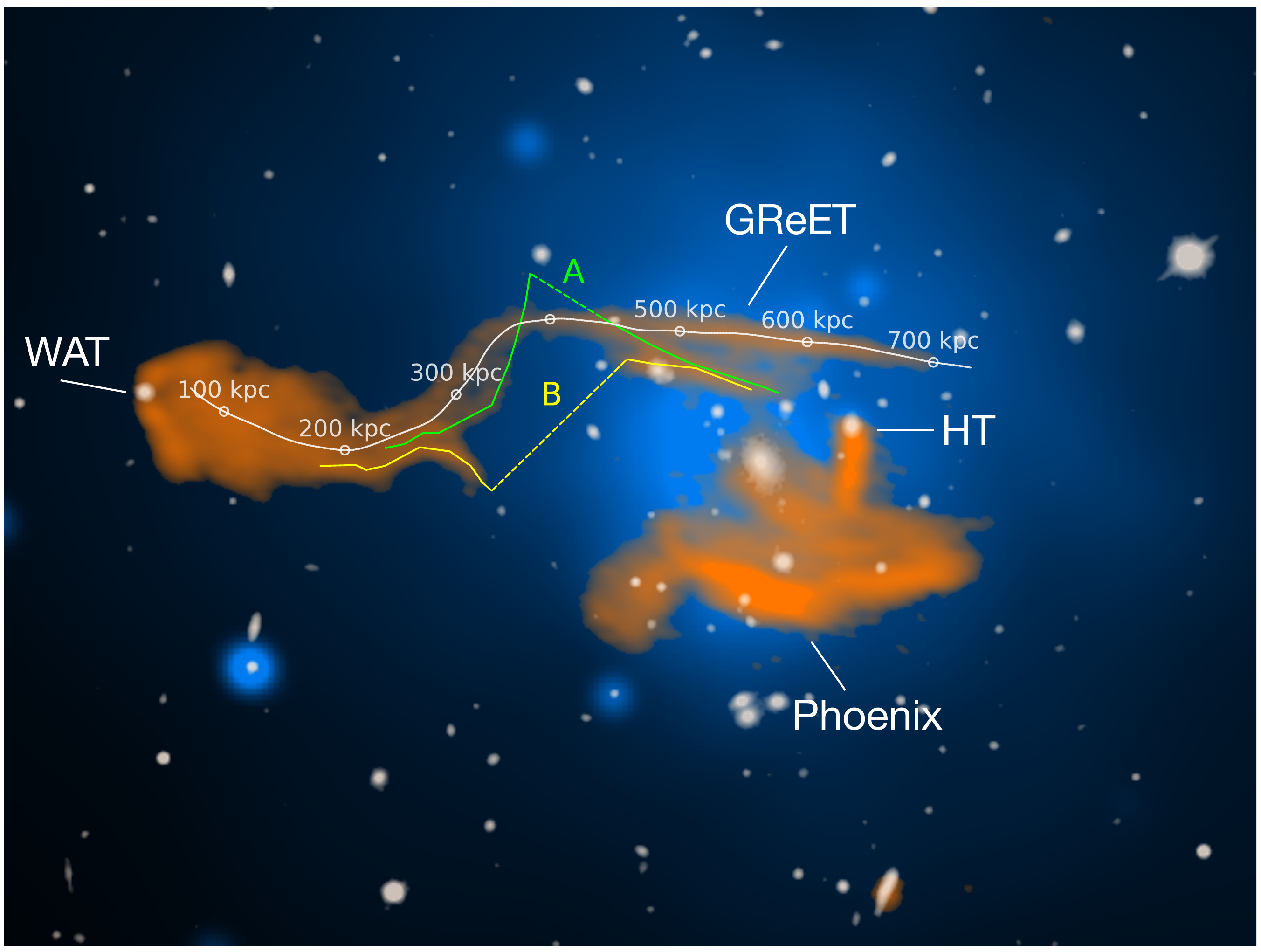}
    \caption{Color-composite of the high-resolution LOFAR HBA (orange), the Chandra X-ray (blue) and the SDSS g \citep[white,][]{Alam2015} images. The white line corresponds to the path that is used to analyze the spectral properties of the GReET, and the dots mark the distance from the starting point in increments of 100\,kpc. Paths \textit{A} (green) and \textit{B} (yellow) highlight the location of possible interrupted structures. The dashed section shows the discontinuity.}
    \label{fig:overview}
\end{figure}

\subsection{LOFAR HBA data}

A1033 was covered by the pointings P155+27 and P157+35 observed in the context of LoTSS \citep{Shimwell2017,Shimwell2019,Shimwell2022}. The analysis of P157+35 has been presented in \citet{deGasperin2017}. Here, we exploit both datasets, taking advantage of the latest developments in the processing pipelines, to increase the sensitivity of the images toward the cluster.
LoTSS is observing the entire northern sky with 8\,h observations per pointing, covering the frequency range 120-168\,MHz. LoTSS pointings are processed by the LOFAR Surveys Key Science Project using the automated direction-independent and -dependent calibration pipelines described in \citet{vanWeeren2016}, \citet{Williams2016}, \citet{deGasperin2019}, \citet{Tasse2018}, and \citet{Tasse2021}. A1033 is located within the newly released area of LoTSS Data Release 2 \citep[DR2;][]{Shimwell2022}, and because a cluster in the PSZ2 catalog \citep{Planck2016}, it is also included in the sample of PSZ2/LoTSS-DR2 clusters recently analyzed by \citet{Botteon2022}. Here, we used the reprocessed dataset presented in \citet{Botteon2022}, in which the data quality has been improved using the extraction and self-calibration procedure described in \citet{vanWeeren2021}. We used \texttt{WSCLEAN} \citep{Offringa2014,Offringa2016} to reimage the data and study the properties of the diffuse emission in the cluster. The final HBA image is shown in \autoref{fig:imgc}. It has a resolution of $12\times6,$ and with a noise level of $110\,\mathrm{{\mu}Jy}\,\mathrm{beam}^{-1}$, it is about 40\% deeper than the image previously presented in \citet{deGasperin2017}.

\section{Results}
In \autoref{tab:images} we summarize the medium- and high-resolution LBA as well as the HBA and GMRT images of A1033 displayed in \autoref{fig:images}. In \autoref{fig:overview}, the radio sources in the cluster are labeled. A connection between the WAT radio galaxy east of the cluster and the GReET is visible at 54 and 144\,MHz. A second radio galaxy with a head-tail morphology is marked  HT in the figure.

For our analysis, we assumed a systematic flux scale uncertainty of 10\% for the LOFAR LBA, LOFAR HBA, and the GMRT \citep{deGasperin2021,Shimwell2022,Chandra2004}.
For all radio images, a common lower $uv$-cut of $100\,\lambda$ was used during imaging, and all imaging was carried out using \texttt{WSCLEAN}.
When information from multiple frequencies is combined for spectral index mapping or spectral shape-modeling, a number of preprocessing steps are required to ensure that the images are comparable. For each image, a point-source catalog was created using \texttt{PyBDSF}, and the images were aligned by cross-matching point sources to account for potential astrometric errors at a level below 5''. Next, all images were convolved to the smallest common circular beam. Last, they were regridded on the same pixel layout.

\subsection{Spectral index maps}\label{sec:spidx}
\begin{figure*}
\centering
\begin{subfigure}{0.33\textwidth}
    \centering
    \includegraphics[width=1.0\linewidth]{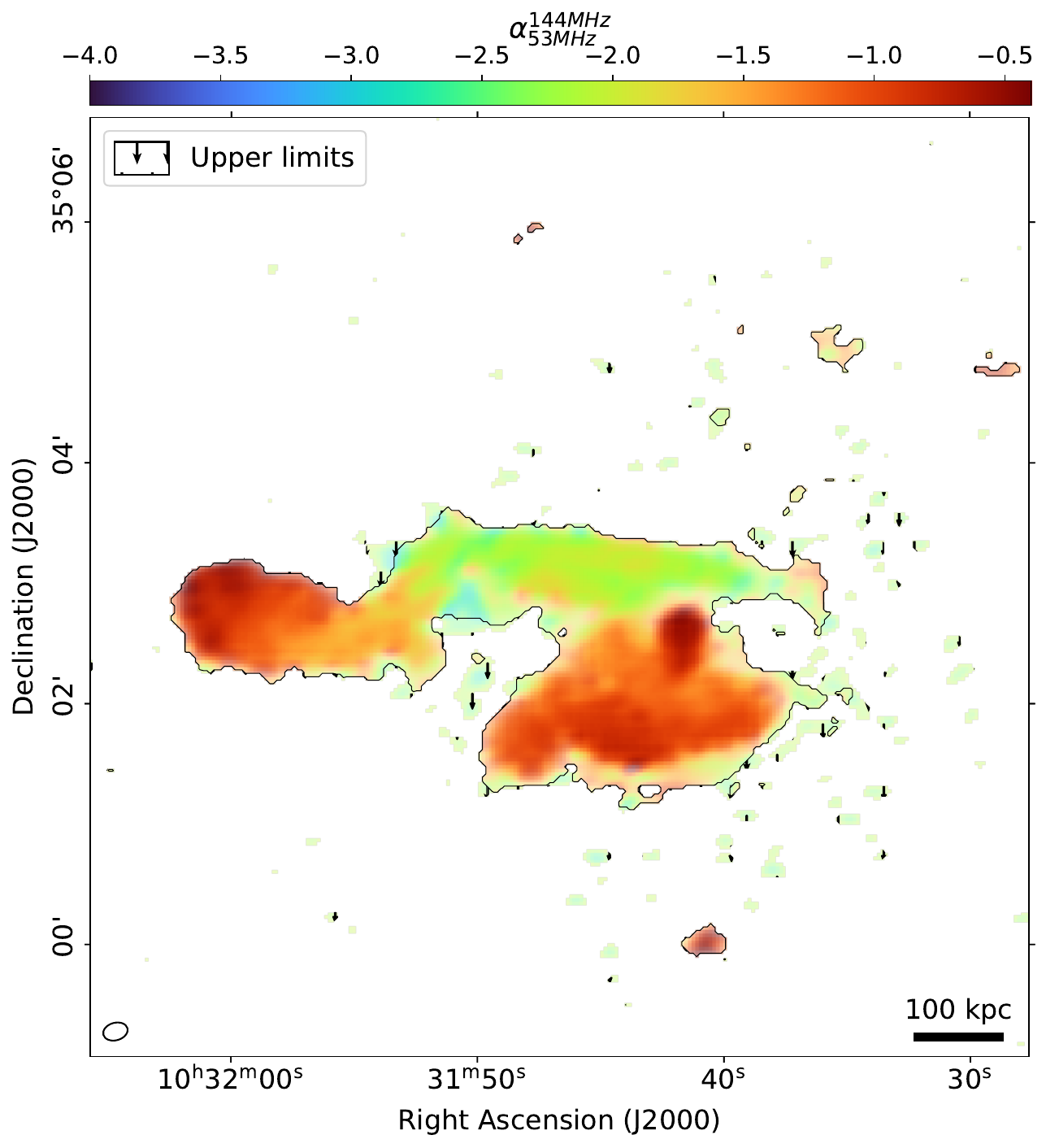}
    \label{fig:silo}
\end{subfigure}
\begin{subfigure}{0.33\textwidth}
    \centering
    \includegraphics[width=1.0\linewidth]{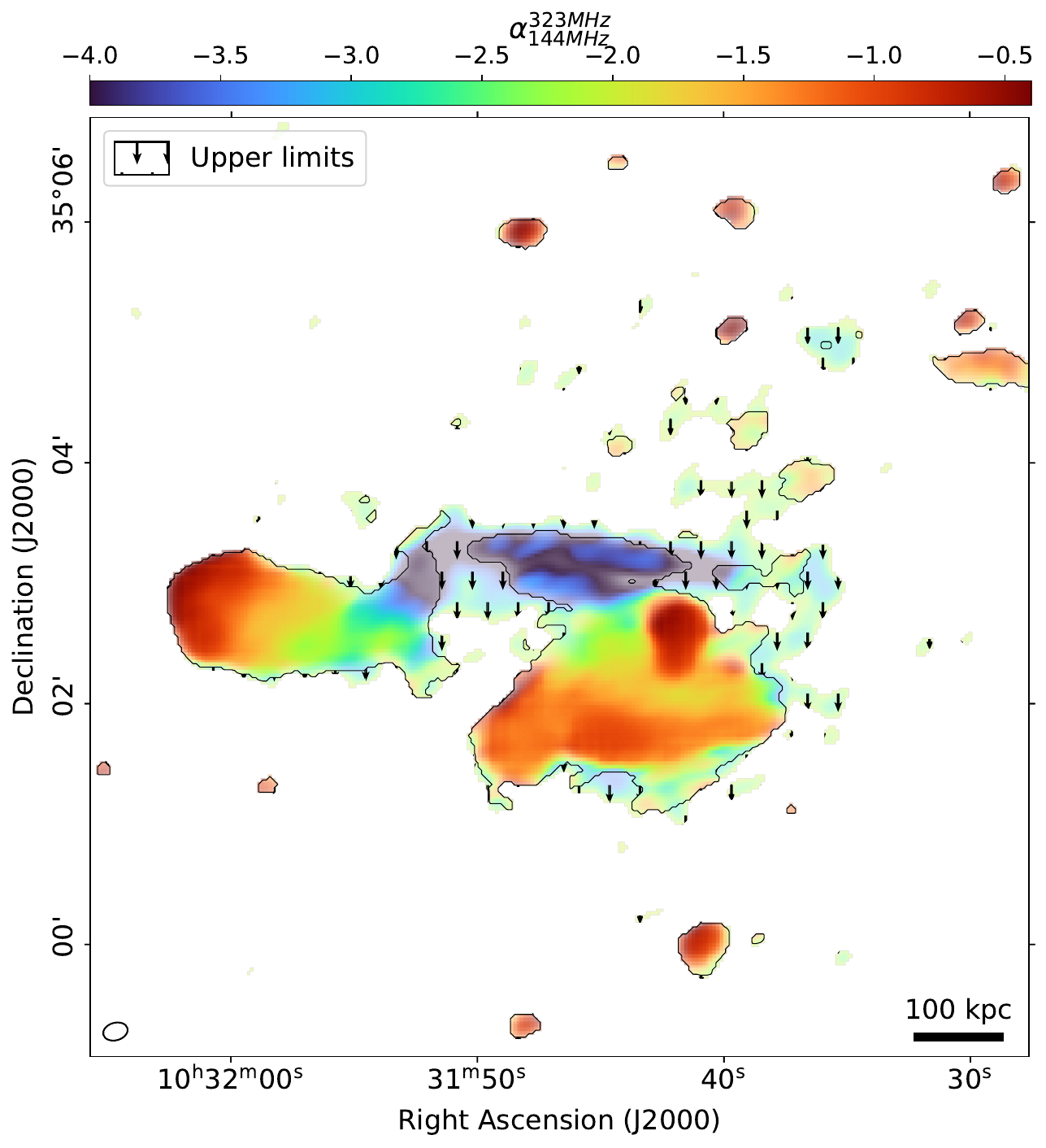}
    \label{fig:sihi}
\end{subfigure}
\begin{subfigure}{0.33\textwidth}
    \centering
    \includegraphics[width=1.0\linewidth]{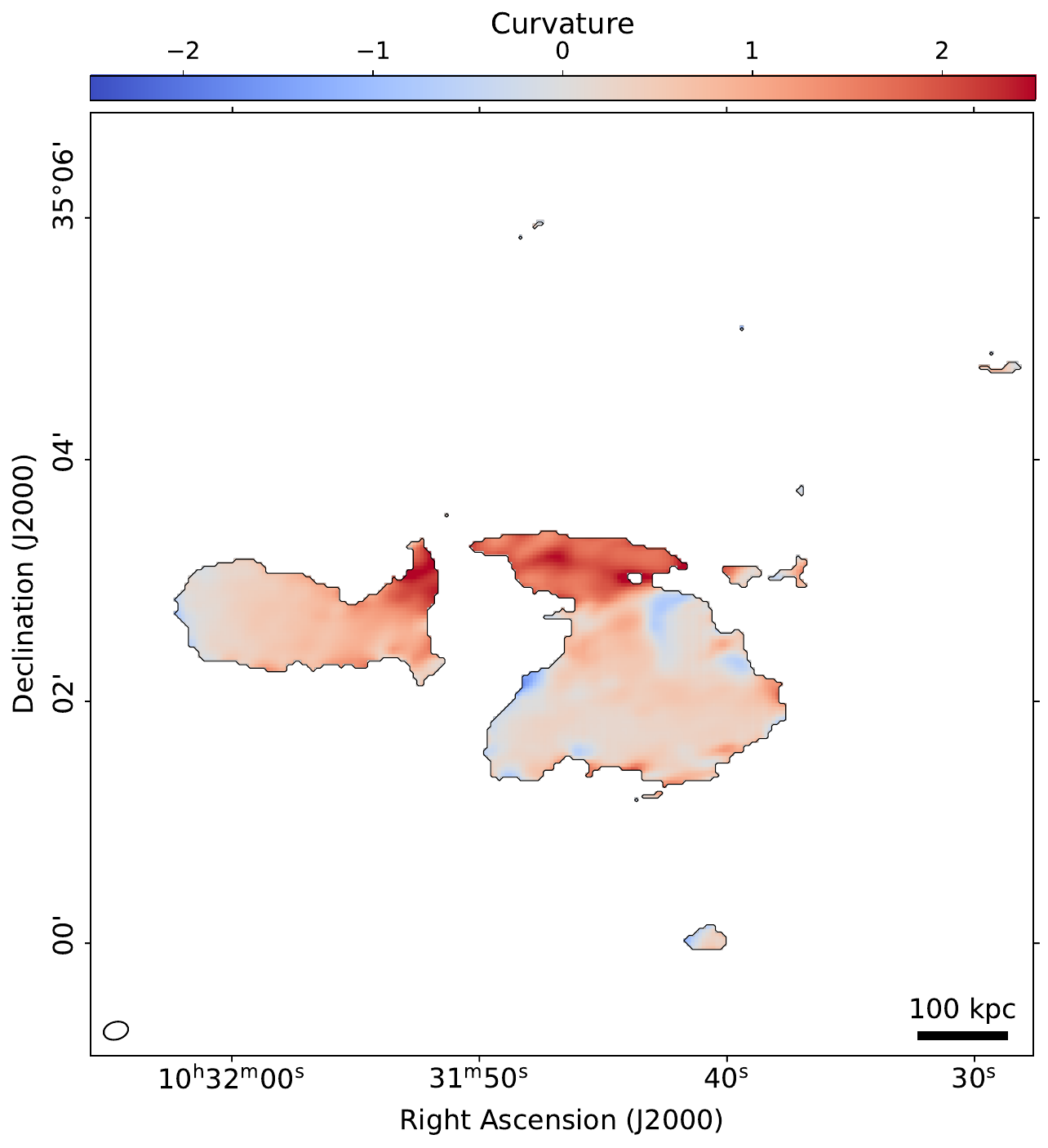}
    \label{fig:curv}
\end{subfigure}
\raggedright
\caption{High-resolution spectral index map of A1033. The left image shows the spectral index between 54 and 144\,MHz, and the right image shows the same between 144 and 323\,MHz (center). The images are on the same color scale and have an angular resolution of $12''\times 9''$. Upper limits below -1.6 are outside the black contours and marked by arrows. The saturation scales inversely with the uncertainty on the spectral index. The right image shows the corresponding curvature map, defined as the difference $\Delta_\alpha = \alpha_{54}^{144} - \alpha_{144}^{325}$.}\label{fig:si}
\end{figure*}
We created spectral index maps in the ranges 54-144\,MHz and 144-323\,MHz. For this, we discarded all pixels with a surface brightness significance below $3\sigma_{\mathrm{rms}}$ in at least one of the two relevant images. We used the linear least-squares method in log-space to determine the spectral index values. The uncertainties were estimated according to Gaussian propagation of uncertainty. To investigate the spectral shape, we created a curvature map from the spectral index maps. The curvature is defined as $\Delta_{\alpha}=\alpha_{54}^{144} - \alpha_{144}^{325}$. The high-resolution  spectral index and curvature maps are displayed in \autoref{fig:si}, and the corresponding uncertainty maps are provided in \autoref{fig:sierr}. 
At the position of the WAT host galaxy, we measure a spectral index of $\alpha=-0.65$ in both frequency intervals. Along the tail, the spectral index gradually steepens. The steepening is stronger at higher frequency. At the location of the bend in the tail, the spectral index does not steepen further and remains approximately constant along the GReET for both frequency ranges. For this section of the tail, we measure an extremely steep spectral index of $-3.8\pm0.18$ between 144 and 325\,MHz,  which is fairly uniform for a projected distance of 300\,kpc. Toward ultra-low frequencies, the spectrum of this region shows extreme curvature ($\Delta_\alpha =1.8\pm0.23$). 

\subsection{Spectral aging analysis}\label{sec:aging}

\begin{figure}
    \centering
    \includegraphics[width=\linewidth]{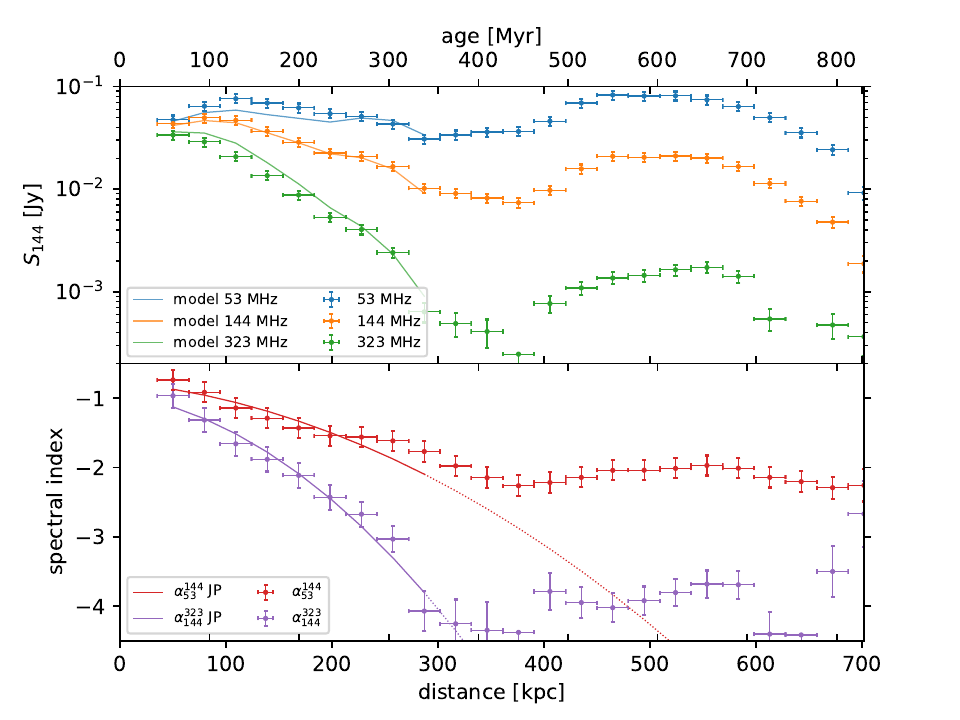}
    \caption{Flux density (top) and spectral index (bottom) measured in the 12.5'' beam-sized regions shown in \autoref{fig:overview} as a function of distance and estimated age. The continuous lines show the best-fitting model ($\chi^2/\mathrm{d.o.f.} = 29.5/17$) assuming minimum aging ($B=2.3\,\mathrm{{\mu}G}$) with a projected velocity of 780\,km/s. The dotted lines show an extrapolation of this model.}
    \label{fig:fluxes}
\end{figure}
\begin{figure}
    \centering
    \includegraphics[width=\linewidth]{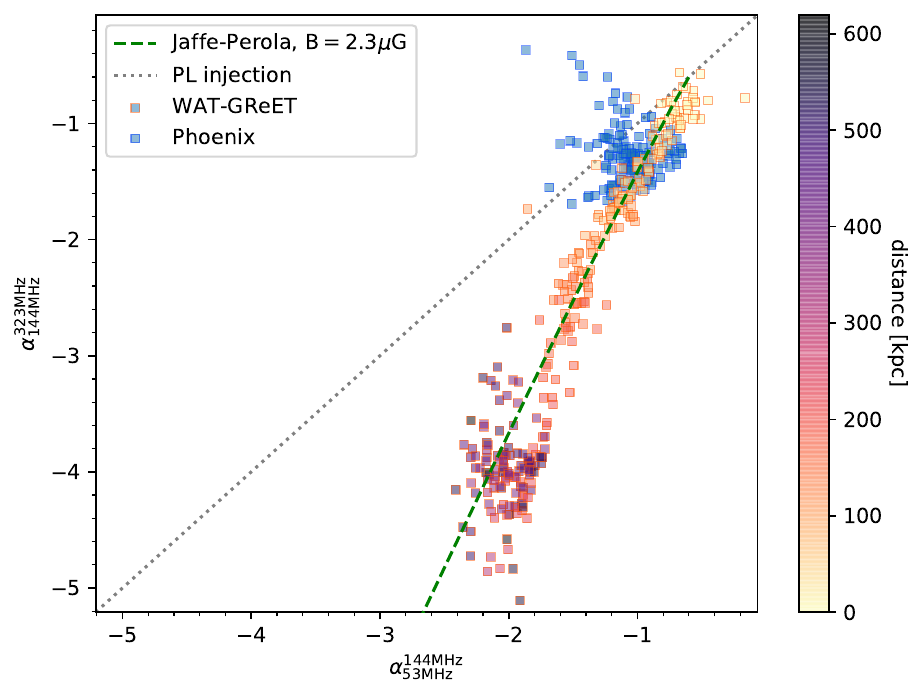}
    \caption{Color-color plot of A1033. Each square represents a $5''\times 5''$ region of the WAT/GReET (orange edges) or phoenix (blue).  For the WAT/GReeT area, the square faces are color-coded according to their approximate distance from the injection point. Additionally, two lines show the spectral properties for a power-law injection (dotted gray) and for a JP aging model with $B=2.3\,\mathrm{{\mu}G}$ (dashed green).}
    \label{fig:colorcolor}
\end{figure}
In absence of acceleration mechanisms, the electron population in a radio galaxy tail is subject to aging due to synchrotron and inverse Compton losses. This causes a spatial evolution of the spectral shape along the tail depending on the magnetic field, injection history, and injection spectral index of the active galactic nucleus (AGN) as well as the projected velocity of the radio galaxy. In the following, we consider this pure-aging scenario for the WAT and examine whether it agrees with our multifrequency observations. This is the expected scenario when no acceleration mechanisms, usually related to ICM shocks, are present in the tail.
For our spectral aging model, we probed a path along the brightest part of the tail where we measured the flux densities at 54, 144, and 323\,MHz and the spectral indices between 54-144 and 144-323\,MHz in 23 different 12.5'' beam-sized regions that were separated by the beam FWHM. This path is shown in \autoref{fig:overview}, and the flux density and spectral index evolution as a function of distance is shown in \autoref{fig:fluxes}. 
We considered the Jaffe-Perola (JP) plasma aging model \citep{Jaffe1973}, which is the most commonly used model for the aging of radio galaxies, to describe the spectral distribution for the first 300\,kpc of the tail, before the brightness of the tail increases again \citep{deGasperin2017}. We performed a nonlinear least-squares fit of the JP model to the spectral index points, assuming that the projected velocity $v_\bot$ of the galaxy along the path is constant.

From the spectral index maps, we estimate the injection index to be $\alpha_{\mathrm{inj}}=-0.65$. The aging history of our model then depends on the magnetic field $B$ and the velocity $v_\bot$. 
Unfortunately, the magnetic field and the projected velocity are strongly correlated, implying that it is not possible to distinguish between a scenario with a magnetic field close to the minimum aging field strength and a lower velocity and a scenario with more rapid losses and a higher velocity of the WAT galaxy (as illustrated in \autoref{fig:merit}). Therefore, we considered a scenario that maximizes the lifetime of the electrons, which provides an upper limit on the radiative age of the plasma. The lifetime of the CRe in the relevant energy regime depends on the magnetic field $B$  as $\tau_\mathrm{age} \propto \sqrt{B} / \left( B^2 + B_\mathrm{CMB}^2 \right)$ \citep{Stroe2014}, where $B_\text{CMB} = 3.2\times (1+z)^2\,\mathrm{{\mu}G}$ \citep{Longair2011} is the magnetic field strength equivalent to the cosmic microwave background (CMB) energy density at redshift $z$. The expression for $\tau_\mathrm{age}$ takes its maximum value at the minimum loss magnetic field $B=B_\mathrm{min}$, where the loss rates due to synchrotron radiation and inverse Compton scattering with the CMB are equal. This is given by $B_\text{min}=B_\text{CMB}/\sqrt{3}$ at the redshift of A1033, $B_\text{min}= 2.3$\,$\mu$G.
We used the nonlinear least-squares method to fit a JP model to the spectral index data points (see \autoref{sec:model}).
By fitting the model to the spectral index points and not the flux densities, we can eliminate the normalization factors and thereby reduce dependences on the injection history of the AGN. 
The best-fitting model has a projected velocity of $v_\bot = 780$\,km/s with a reduced $\chi^2$ statistic of $\chi^2/\mathrm{d.o.f.} = 29.5/17$, which agrees reasonably well with the $\approx 730\,$ km/s found in \citet{deGasperin2017}. This corresponds to a maximum age of the GReET of 800\,Myr. The spectral index and flux density data points as well as the best-fitting model are shown in \autoref{fig:fluxes}. For the flux density models, the flux normalization was fit in log-space independently as a free parameter for each beam-sized region.

To complement the spectral analysis, we created a color-color map. To do this, we calculated the 54-144 and 144-323\,MHz spectral indices in 5'' square boxes in a region covering the WAT and GReET and a second region covering the radio phoenix. We considered all boxes within the two regions with a significance above $3\sigma_\mathrm{rms}$ at all three frequencies. For comparison, we also show the spectral trajectory of a power-law injection and the JP aging model discussed above. For points along the WAT and GReET, we additionally color-code the distance from the injection point along the tail. The resulting diagram is shown in \autoref{fig:colorcolor}. The boxes associated with the radio phoenix occupy a rather confined region in the spectral index space without clear trends. In contrast, along the WAT, there is a clear steepening with increased distance from the injection point. The spectral index points departure from the PL-injection spectrum along a trajectory that agrees with a JP model with increased distance. After ~300\,kpc, the steepening between 54-144 and 144-323\,MHz stops and the boxes corresponding to the following part of the GReET scatter around $\alpha_{144}^{323}=-4$ and $\alpha_{54}^{144}=-2$. 

\subsection{Source-subtracted low-resolution images} \label{sec:sub}
\begin{figure*}
\centering
\begin{subfigure}{0.33\textwidth}
    \centering
    \includegraphics[width=1.0\textwidth]{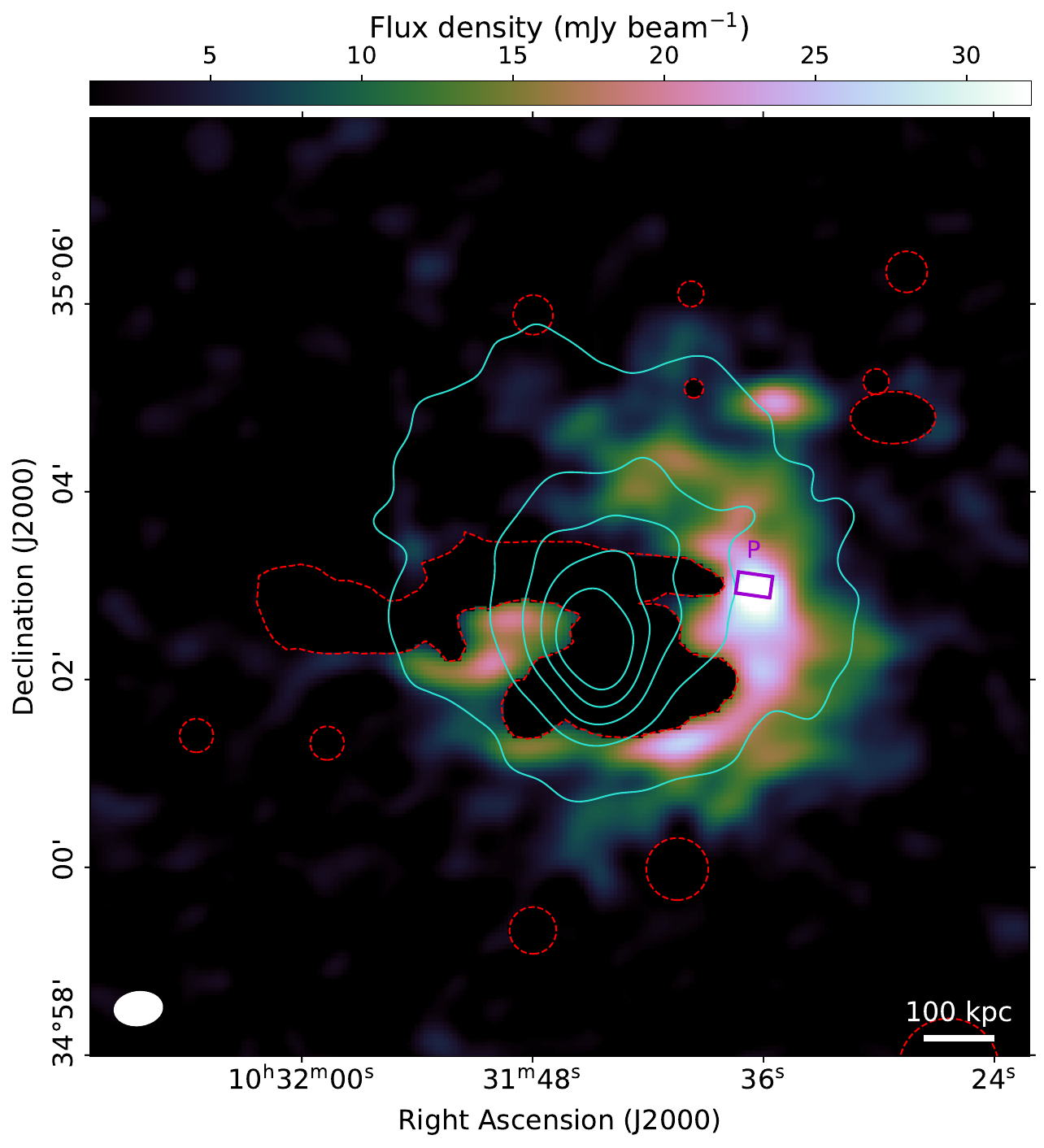}
    \caption{54\,MHz}
    \label{fig:imghalolba}
\end{subfigure}
\begin{subfigure}{0.33\textwidth}
    \centering
    \includegraphics[width=1.0\textwidth]{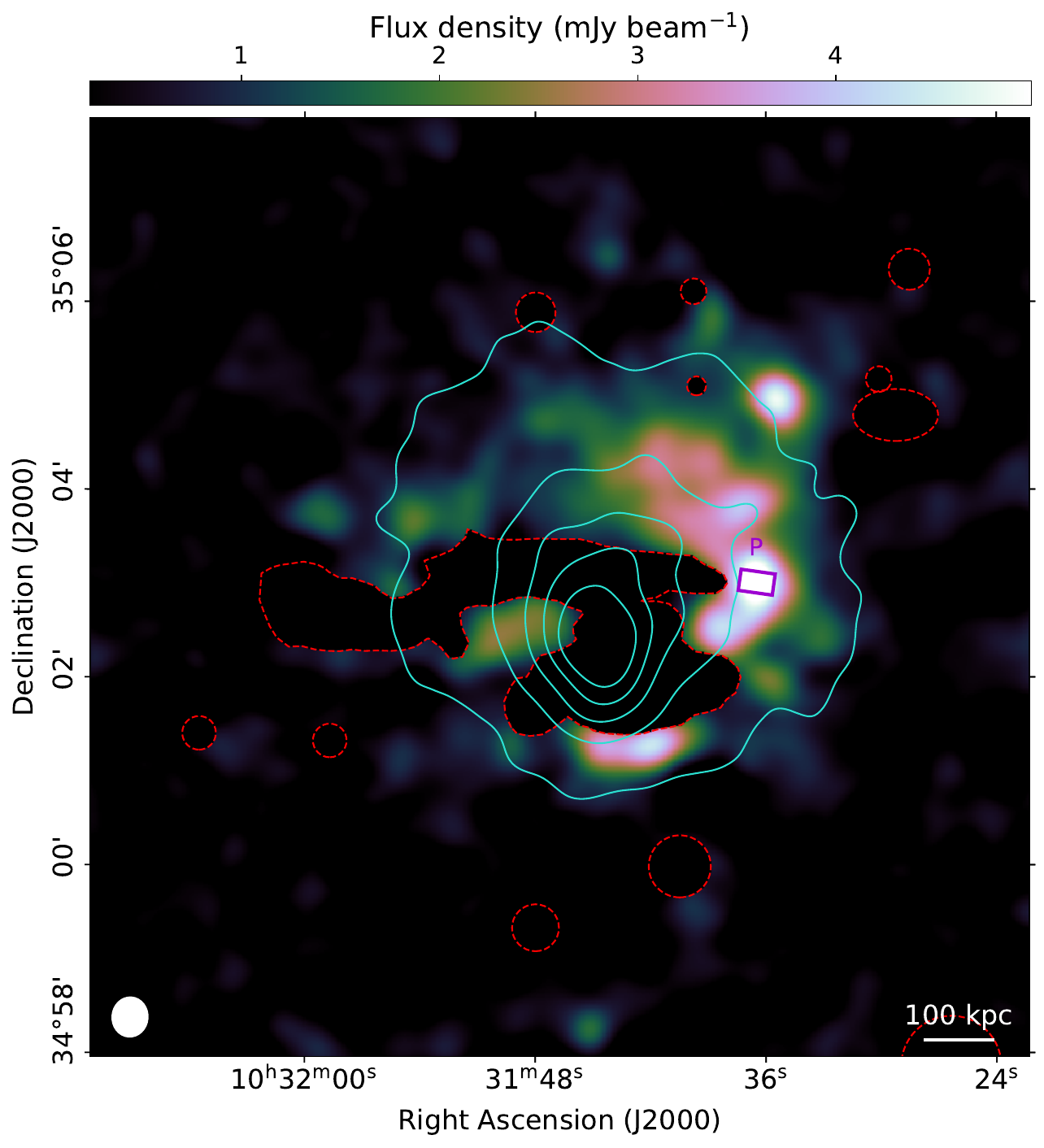}
    \caption{144\,MHZ}
    \label{fig:imghalohba}
\end{subfigure}
\begin{subfigure}{0.33\textwidth}
    \centering
    \includegraphics[width=1.0\textwidth]{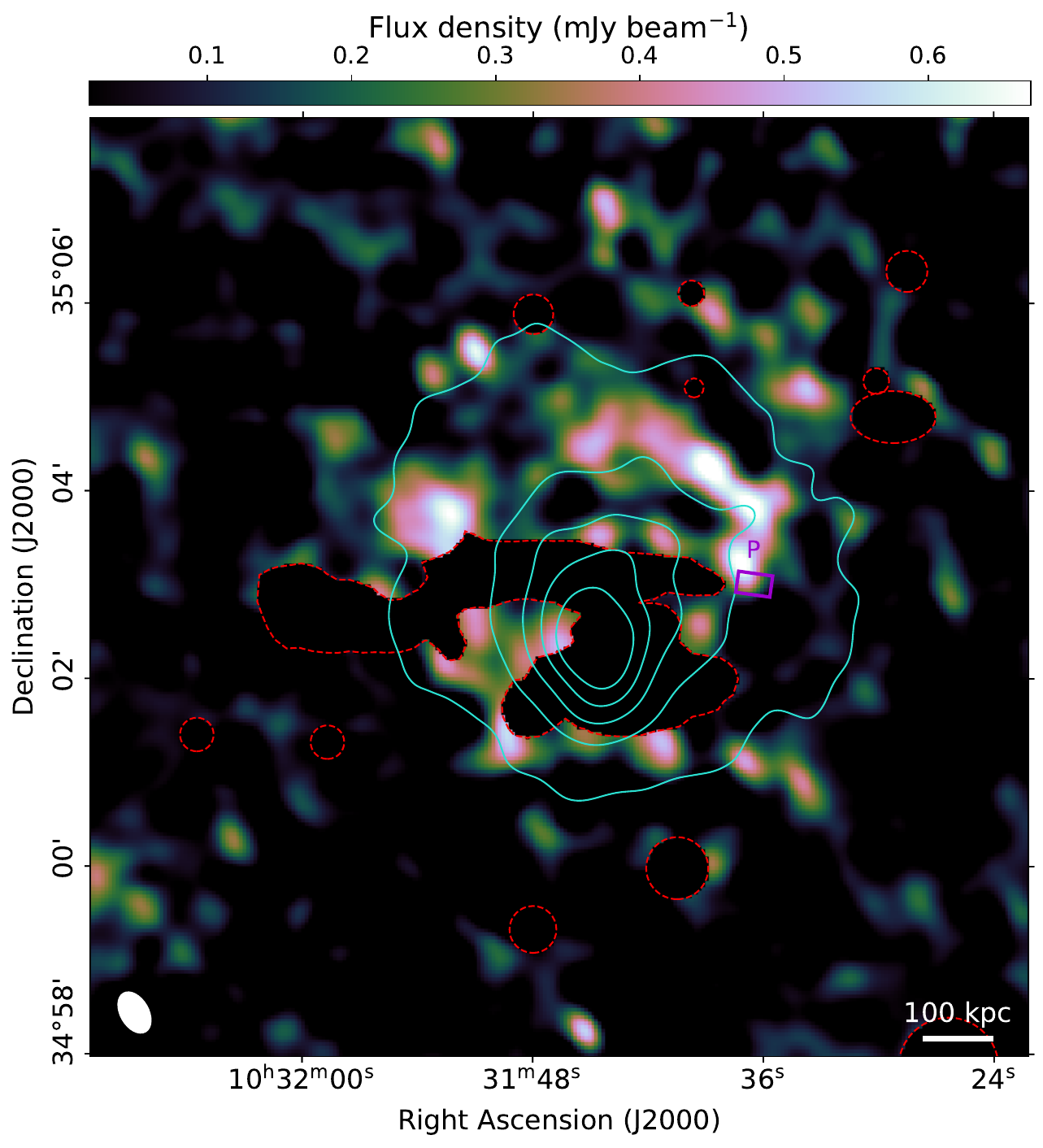}
    \caption{325\,MHZ}
    \label{fig:imghalogmrt}
\end{subfigure}
\caption{Source-subtracted low-resolution images. The image weights are tapered to a resolution of 25''. Red contours indicate sources that were subtracted from the data, and cyan contours are smoothed Chandra X-ray contours in the energy band 0.5-7\,keV \citep{deGasperin2015}, linearly spaced between 10-100$\sigma$, where $1\sigma = 6.6\times10^{-6}$\,photons\,s$^{-1}$\,cm$^{-2}$\,arcmin$^{-2}$. The purple rectangle labeled \textit{P} marks the brightest part of the visible halo.}\label{fig:imghalo}
\end{figure*}
The medium-resolution images \autoref{fig:imga} and \autoref{fig:imgc} show an extended diffuse radio emission superimposed on the bright sources associated with individual radio galaxies, the GReET and the radio phoenix.  In the LOFAR images, emission with an extension of several 100\,kpc is visible at 54\,MHz as well as at 144\,MHz. This emission is associated with the radio halo detected in A1033 \citep{Botteon2022}.
To isolate the emission of the radio halo from the foreground sources, we obtained a model for the compact sources based on high-resolution (Briggs~$=-1.0$) images at 54\,MHz, 144\,MHz, and 323\,MHz. This model is restricted to the clean components in a region that only includes the sources in the high-resolution images (e.g., the radio galaxies, GReET, and phoenix), and subtracted from the data, identical regions are used at all three frequencies.  We reimaged the subtracted data using baselines in the range 100-20000$\,\lambda$ and tapering the imaging weights to a resolution of 25'' to enhance the extended emission. The resulting images are displayed in \autoref{fig:imghalo} together with the regions we used to restrict our subtraction- model. 
In the LOFAR maps at 54 and 144\,MHz in \autoref{fig:imghalo}, the extended emission is significantly detected at a surface brightness above $5 \sigma_\mathrm{rms}$ for a region with a diameter of $\approx700\,$kpc except for the subtracted regions. The shape is approximately circular in both frequencies, but some morphological differences exist. At 54\,MHz, the halo appears to extend further to the southwest, while a hole in the emission is visible in the northeast. 
Given the complex procedure of the calibration of a bright, extended source with complex morphology superimposed on the faint, even more extended halo emission together with the source subtraction, we cannot exclude the possibility that these differences in morphology in the lower-significance regime are not physical, but rather artifacts in the data.
At 323\,MHz, the detected emission appears more patchy, probably because of artifacts caused by calibration errors or radio frequency interference in the GMRT map. In the low-resolution images, we also display the point-source-subtracted Chandra X-ray contours from \citet{deGasperin2015}. The shape of the radio emission is approximately circular and mostly confined within the X-ray contours.

\subsection{Calculating the radio halo flux}
To estimate the total flux density of the halo, we employed the halo flux density calculator \texttt{Halo-FDCA} \citep{Boxelaar2021}, an MCMC code for estimating halo flux densities with reduced human bias. The code can handle masks to exclude image regions from the analysis. This allowed us to exclude the subtracted regions that do not contain reliable emission. We used the code to fit a spherically symmetric exponential surface-brightness profile described by
\begin{equation}
    I(r) = I_0 e^{-\frac{r}{r_e}}
\end{equation}
to the data. Here, $I(r)$ is the surface brightness at a distance $r$ from the profile center, $I_0$ is the peak surface brightness, and $r_e$ is the \textit{e}-folding radius \citep{Murgia2009}. This empirical model has frequently been found to describe radio halos accurately \citep[e.g.,][]{Vacca2014,Osinga2021,Botteon2021,Botteon2021b} and it comes with a minimum number of free parameters, which is beneficial for the halo in A1033, which is partly obscured by other sources.   
A similar analysis for A1033 in LOFAR HBA is presented in \citet{Botteon2022}. To allow a consistent analysis of the LBA and HBA data specifically tailored to A1033, we repeated the source subtraction, imaging, and fitting procedures also for the HBA data.

The result of the halo flux calculation is a $1.46\pm0.15$\,Jy at 54\,MHz and $0.29\pm0.03$\,Jy at 144\,MHz. The detailed parameters are listed in \autoref{tab:halofit}, and plots of the halo fit result and fit residual are shown in \autoref{fig:halofit}. The location and extension of the best-fitting models at 54\,MHz and 144\,MHz agree well with \textit{e}-folding radii of $154\pm2$ and $165\pm3$\,kpc, respectively. The LBA model is slightly farther south and the radius is smaller by 7\%. We also manually measured the flux density  within the respective $3\sigma_\text{rms}$ contours $F_{3\sigma}$ of the halo in \autoref{fig:halofit}, obtaining $F^{3\sigma}_{ 54\,\text{MHz}}=0.87\pm0.09$ and $F^{3\sigma}_{ 144\,\text{MHz}}=0.15\pm0.02$. Due to the missing extrapolation of the emission in the masked regions, the manually measured flux is a lower limit for the total flux of the halo. The ratio of the manually measured and the fitted flux density is different between LBA and HBA (0.6 and 0.52). We attribute this to a small difference in the location and size of the fitted surface brightness profile.
We estimate the total radio power at 150\,MHz of the halo based on the \texttt{Halo-FDCA} results. The resulting value is $P_\mathrm{150\,MHz} = 1.22\pm 0.13 \times 10^{25}$\,W\,Hz$^{-1}$, where a spectral index of $\alpha=-1.65$ was used for extrapolation and $k$-correction, as we justify in the following \autoref{sec:halosi}. The independent estimate of the radio power in \citet{Botteon2022} ($P_{150} = 1.09\pm 0.28 \times 10^{25}$\,W\,Hz$^{-1}$) is consistent with ours (at about $1\sigma$), but their reconstructed morphology differs significantly ($r_e=165\pm3$ to $r_e=103\pm1$). We interpret this to mean that the halo morphology is more susceptible to differences in masking and fitting than the flux density of the halo. In general, the fitted parameters of the halo model are subject to a systematic uncertainty connected to the masking and subtraction procedure, which is hard to assess and was therefore not accounted for in our uncertainties.
\begin{table*}[h]
    \centering
    \caption{Halo fit results. The total flux density $F_\text{FDCA}$ includes an extrapolation of the model in the masked region.}
    \begin{tabular}{c c c c c c}
         $\nu$ [MHz] & $I_0$ [$\mu$Jy$/$arcsec$^{2}$] & $r_e$ [kpc] & $F_{\text{FDCA}}$ [Jy] &$F_{3\sigma}$ [Jy] &$P_\nu$ [$10^{25}$W/Hz] \\\hline
        54 & $50.33\pm5.14$ & $154\pm2$ & $1.46\pm0.15$ & $0.87\pm0.09$ & $6.56\pm 0.67$ \\ 
        144 & $8.63\pm0.90$ & $165\pm3$ & $0.29\pm0.03$ & $0.15\pm0.02$ & $1.30\pm0.13$ \\
    \end{tabular}
    \label{tab:halofit}
\end{table*}

\subsection{Spectral index of the radio halo}\label{sec:halosi}
\begin{figure}
    \centering
    \begin{subfigure}{0.8\linewidth}
    \includegraphics[width=1.0\linewidth]{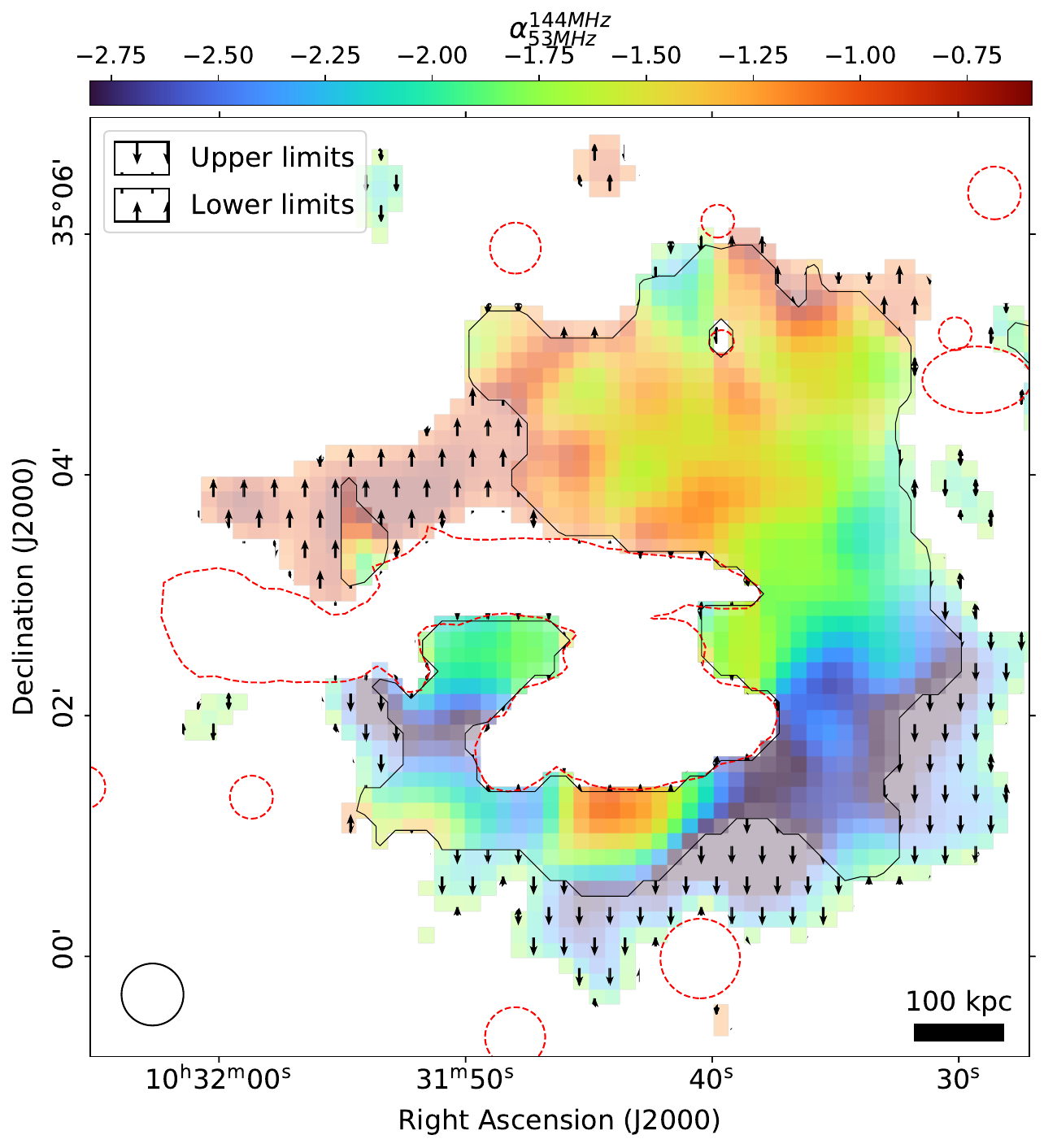}
    \end{subfigure}
    \begin{subfigure}{0.8\linewidth}
    \includegraphics[width=1.0\linewidth]{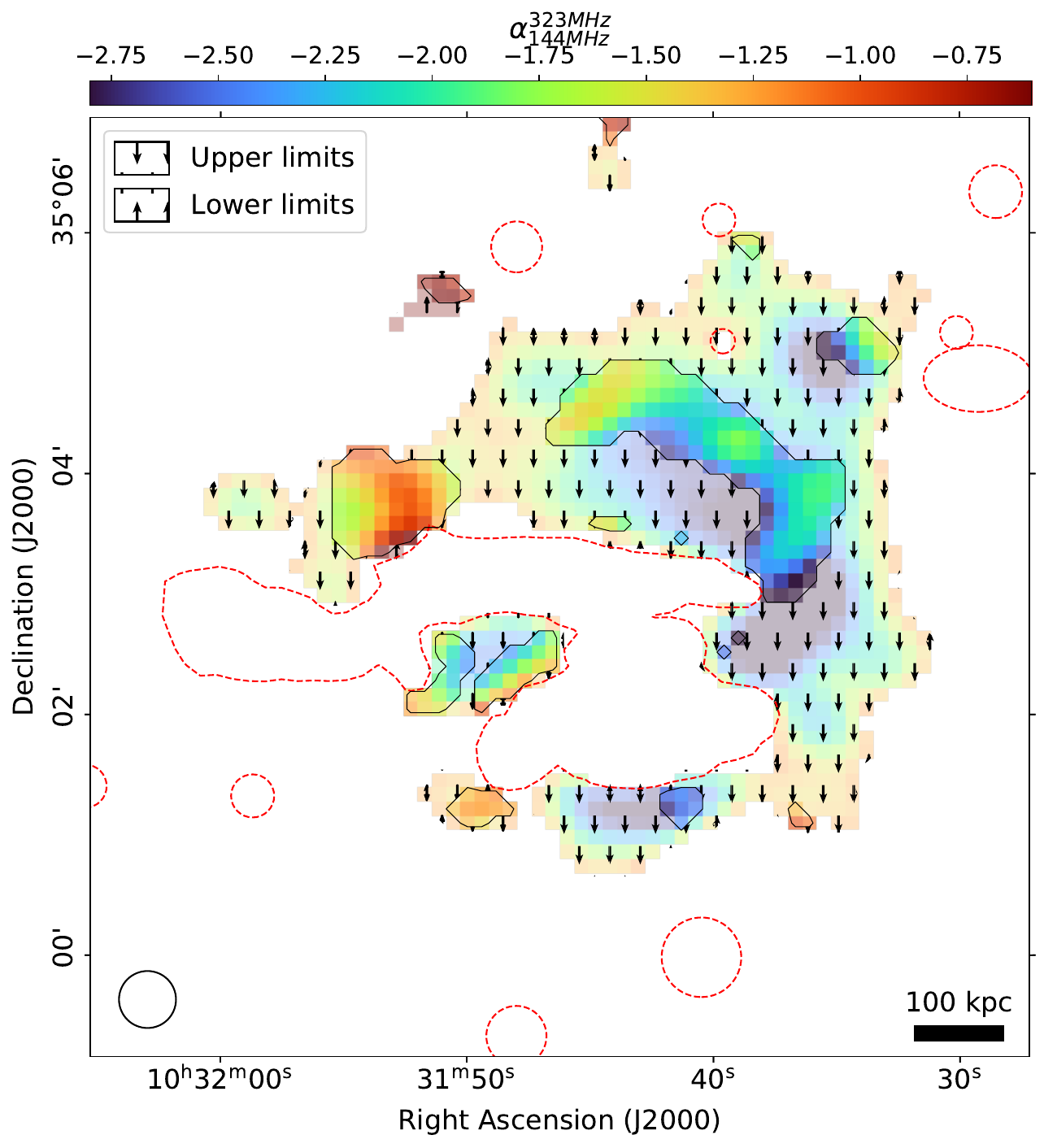}
    \end{subfigure}
    \caption{Source-subtracted spectral index maps at a spatial resolution of 30''. The top panel shows the spectral index between 54\,MHz and 144\,MHz, and the bottom panel shows the same between 144\,MHz and 325\,MHz. The $3 \sigma$ upper and lower limits are indicated by arrows. Dashed red lines highlight the regions from which sources were subtracted. }\label{fig:sisubtract}
\end{figure}
We used the source-subtracted images presented in \autoref{fig:imghalo} to create maps of the low-resolution spectral index and to study the spatial variation of the spectral index within the radio halo. The spectral index maps were created as described in \autoref{sec:spidx}. We excluded the regions that were used to subtract the compact sources. These maps are displayed in \autoref{fig:sisubtract}, and the uncertainty maps are shown in \autoref{fig:sisubtracterr}. In the range of 54-144\,MHz, the spectral index of the halo could be determined significantly across a large area. The northern part of the halo is less steep with a spectral index of about $\alpha=-1.5$, and the southwestern region is steeper than $\alpha=-2$. The same trend extends to the ultra-steep upper limits in the southwest and to the flat-spectrum features expressed by lower limits in the northeast. As stated in \autoref{sec:sub}, we cannot exclude the possibility that these features are at least partly a consequence of systematic errors in the calibration. Because only a part of the radio halo is detected above a significance of $3\sigma$ in the 323\,MHz GMRT map, we can constrain the spectral index in the range 144-323\,MHz only in a comparatively small area. In this region northwest of the GReET, the spectral index is $\alpha_{144}^{323}= -2.15\pm 0.19$. For the majority of the remaining halo area, we can constrain the spectrum to be steeper than $\alpha=-2$. This suggests that the spectrum of the halo is curved.

To estimate the integrated halo spectral index between 54 and 144\,MHz, we compared different methods. Using the halo fits, we derive a spectral index  of $\alpha_{\text{FDCA}}=-1.65\pm0.17$ based on the integrated surface brightness. If the flux density is measured manually in the region where we have $>5\sigma$ significance at both frequencies, the corresponding integrated spectral index is $\alpha_{\mathrm{man}}^{ 5\sigma}=-1.74\pm0.14$. A more conservative estimation using only the area above $10\sigma_{\mathrm{rms}}$ in both images yields $\alpha_{\mathrm{man}}^{10\sigma}=-1.69\pm0.15$. We adopted the value based on the \texttt{Halo-FDCA} fluxes as reference value for the halo because it does not depend on the sensitivity of the observations.

\section{Discussion}
\subsection{Spectral study of the GReET}\label{sec:sigreet}

In \autoref{fig:fluxes} we show the flux density and spectral index evolution along the tail together with the best-fitting JP model as described in \autoref{sec:aging}.  
After about 200\,kpc from the injection point, the spectral index appears to be systematically flatter than the best-fitting model. To some degree, this deviation might be caused by a change in velocity of the WAT. Presumably, the galaxy reached a maximum velocity at the minimum of its trajectory in the cluster potential well and slowed down afterward. This is not accounted for in our constant-velocity assumption.
However, at about $\approx\,350\,$kpc form the injection point, the spectrum does not steepen further and a plateau in spectral index associated with the GReET is reached. This constant and even flattening spectral index cannot be explained with the extrapolation of a pure aging model, even when a change in the WAT velocity is considered. The same stalling in the spectral index trend can also be observed in the color-color diagram in \autoref{fig:colorcolor}, where after initial JP-like aging, the spectral evolution stops and appears frozen for several 100\,kpc. Consequently, an additional energetization mechanism is required. This mechanism needs to be just efficient enough to compensate for the radiation losses, but not so strong that the spectral shape is significantly flattened. 

This was previously concluded in \citet{deGasperin2017}, where turbulent reacceleration of the electrons in the tail by magnetic pumping was proposed as a possible mechanism. The suggested scenario is that turbulence is forced into the tail by interactions with the surrounding ICM, which can provide mild acceleration that acts for sufficiently long timescales to explain the homogeneous nature of the GReET. The extreme curvature we detect fully supports the scenario outlined in the previous work: it is a consequence of a cutoff in the electron energy spectrum caused by the balance between the very gentle acceleration and the cooling of CRe, with an acceleration time of the radio-emitting electrons comparable to their cooling time \citep[e.g.,][]{Brunetti2014}.

\subsection{Turbulent velocity}
In the high-resolution map shown in \autoref{fig:overview}, the tail of the WAT shows disrupted features. We suggest that these can be attributed to large-scale shear flows and turbulent motions in the ICM. The tail of the WAT appears to be split into three filaments after about 250\,kpc from the injection point. The brightest filament directly connects to the GReET, and the fainter filaments, labeled in \autoref{fig:overview} as A and B, are discontinuous. Because the GReET is clearly bifurcated, where one of the two parts has no clear connection to the WAT, we consider the possibility that a shear motion in the ICM, at a scale close to the injection scale of the turbulence, caused either A or B to disconnect from the southern segment of the GReET. 
This then allows us to constrain the large-scale turbulent velocity in the ICM: 
In the diffusion regime, CRs are displaced a distance $d$ after a time $t$ according to $d\sim \sqrt{t D}$, where $D$ is the spatial diffusion coefficient. For super-Alfv\'enic turbulence, we can approximate $D \sim v_L L,$  where $v_L$ is the turbulent velocity at scale $L$. If the displacement $d \sim L$, the velocity of large-scale motions can be estimated as $v_L \sim d/t$ , where $t$ can be constrained from the age of the plasma from our limit on the velocity of the WAT. We estimate the age at the discontinuities of path A and B to be $t_\mathrm{A} < 380$\,Myr and $t_\mathrm{B} < 310$\,Myr using the distance of the discontinuities A and B from the injection point (300\,kpc and 250\,kpc, respectively) as well as our limit on the WAT velocity of $\approx 780\, \mathrm{km\,s^{-1}}$. The corresponding projected length scales of the discontinuity are $d_\mathrm{A} = 65$\,kpc and $d_\mathrm{B} = 144$\,kpc. This means that we can constrain the turbulent velocity to $v_\mathrm{A} \geq 167\,\mathrm{km}\,\mathrm{s}^{-1}$ and $v_\mathrm{B} \geq 464\,\mathrm{km}\,\mathrm{s}^{-1}$. These values fall within the typical range found for large-scale turbulent motions in simulations of cluster mergers, which are several $100\,\mathrm{km}\,\mathrm{s}^{-1}$ \citep{Vazza2018}.
This velocity can also be compared to the sound speed $c_\mathrm{s}$ derived from the ICM temperature $kT_\mathrm{ICM} = 6.15 ^{+1.14}_{-0.88}$\,keV found in the Chandra X-ray analysis of \citet{deGasperin2015}, which yields
\begin{equation}
c_\mathrm{s} = 1480 \left( \frac{T_\mathrm{ICM}}{  10^8 \, \mathrm{K}} \right)^{1/2} \, \mathrm{km}\,\mathrm{s}^{-1} = 1250^{+111}_{-92}\,\mathrm{km\,s}^{-1}.
\end{equation} 
This implies that the turbulent motions estimated above are likely subsonic, in agreement with expectations in the ICM \citep[e.g.,][]{Brunetti2014}.

This consideration raises the question of how the tail can resist turbulent diffusion for a timescale of several $100\,\mathrm{Myr}$. When we assume a Kolmogorov cascade, the turbulent velocity at a scale $L$ depends on the size of the scale as $v_L\propto L^{1/3}$. Thus, at a scale equal to the thickness of the filaments, the velocity derived from our estimate should be $v_\mathrm{A}({10\,\mathrm{kpc}}) \geq 89\,\mathrm{km\,}\mathrm{s}^{-1}$ or $v_\mathrm{B}({10\,\mathrm{kpc}}) \geq 190\,\mathrm{km\,}\mathrm{s}^{-1}$.
If this value is below the Alfv\'en velocity $v_\text{Alfv\'en}$,  which is typically in the range $\sim 100\,\mathrm{km}\mathrm{s}^{-1}$ \citep{Brunetti2014}, Reynolds and Maxwell stresses of magnetic fields may stabilize the narrow tails. Similar considerations have been put forward to explain the stability of radio filaments in AGN bubbles \citep{Brienza2021}. 
From the Navier-Stokes equation, the term corresponding to matter motions on a scale $k$: $\rho {\dif}v^2 k$ competes against magnetic field line tension on the same scale: $B^2k / (4\pi)$. Thus, for the line tension to dominate, we require
\begin{equation} \label{eq:Breq}
    B \geq v \sqrt{4\pi \rho}.
\end{equation}
For a reference density 100\,kpc form the cluster center (at the projected distance to the GReET) of $\rho \approx 10^{-23}\,\mathrm{kg\,m^{-3}}$, which is a typical value for clusters similar to A1033, and a velocity in the range derived above from $v_\mathrm{A}({10\,\mathrm{kpc}}) \geq 89\,\mathrm{km\,}\mathrm{s}^{-1}$ to $v_\mathrm{B}({10\,\mathrm{kpc}}) \geq 190\,\mathrm{km}\mathrm{s}^{-1}$, a magnetic field of at least 3.1 - 6.7\,$\mathrm{{\mu}G}$ is required. For the upper half of this range, the magnetic field in the GReET would be greater than what is usually considered for magnetic fields in the ICM even for significantly more massive clusters \citep[e.g.,][]{Bonafede2010}, which would imply that it must be seeded from the WAT and sustained for the life time of the GReET. Lower values for the magnetic field in the tail are possible if the distance between the cluster center and the GReET is significantly larger than the projected distance, such that the density $\rho$ entering \autoref{eq:Breq} is lower.

Furthermore, we tried to quantify the conditions under which the electrons can be confined in the GReET for 800\,Myr: In general, diffusion perpendicular to magnetic filaments is small, which allows an efficient trapping of the electrons. However, if the magnetic field in the GReET is turbulent, that is, sub-Alfv\'enic as we require to explain the stability, perpendicular diffusion of electrons may be driven by stochastic diffusion of field lines. The perpendicular diffusion length of the CRe $l_{\bot, \mathrm{cr}}^2$ after a time $t$ can be estimated as \citep{Lazarian2014}
\begin{equation}
    l_{\bot, \mathrm{cr}}^2 \sim  \frac{(D t)^{3/2}}{27 L} M_A^4,
\end{equation}
where in this case, the Alfv\'enic Mach-number $M_A\sim1$ and the injection scale $L \sim L_\bot = 10\,\mathrm{kpc}$ is the traverse scale of the GReET.
We define the lifetime of the electrons in the GReET as the time when the perpendicular diffusion is comparable to the traverse size $l_{\bot, \mathrm{cr}} \sim L_\bot$,
\begin{equation}
    t \sim 9 L_\bot^2 / D.
\end{equation}
To be stable for 800\,Myr, a diffusion coefficient below $D= 3.4\times 10^{29}\,\mathrm{cm^2 s^{-1}}$ is required, which agrees well
with the previous estimate of \citet{deGasperin2017} and observational results for other clusters based on studies of the metallicity profiles \citep{Rebusco2005,Rebusco2006}.

\subsection{Radio halo} 

\begin{figure}
    \centering
    \includegraphics[width=1.0\linewidth]{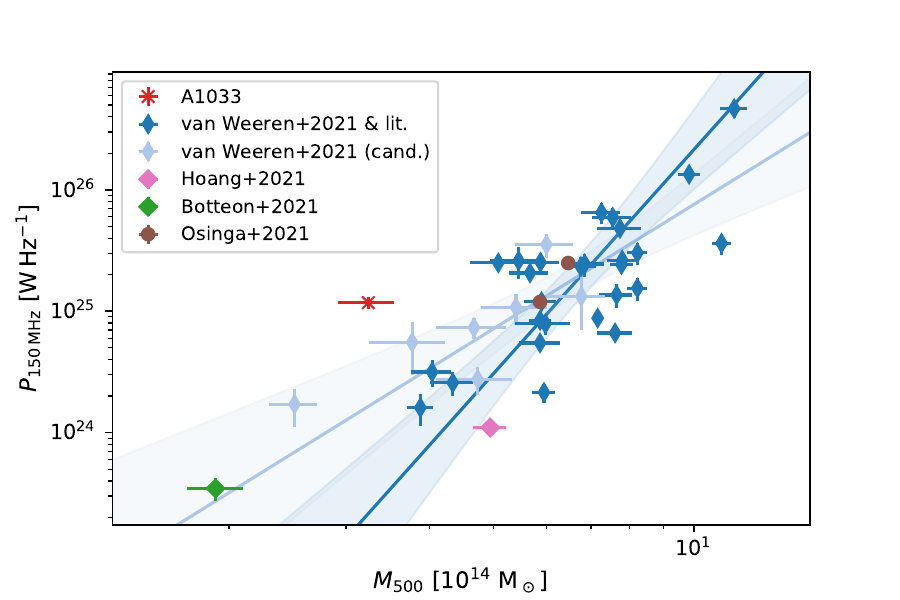}
    \caption{Radio power at 150\,MHz  of low-frequency radio halos as a function of the PSZ2 mass. The lines indicate the extreme cases of the correlations of radio power to cluster mass reported in \citet{vanWeeren2021}, which were derived using different fitting methods and samples. The dark blue line shows the steepest (orthogonal method) and the light blue line shows the flattest (including candidate halos, Y|X method) correlation. The shaded regions show the 95\% confidence interval. The halos reported as `van\,Weeren+21\,\&\,lit.' include a literature radio halo sample \citep{deGasperin2015,Cordey1985,Wilber2018,Botteon2019,Birzan2019,Wilber2018,Savini2018,Savini2019,Bonafede2018,Hoang2019a,Hoang2019b,Macario2013,George2017,Duchesne2021}.}
    \label{fig:radiopower}
\end{figure}

The spectral index of  $\alpha=-1.65\pm0.17$ between 54 and 144\,MHz places the radio halo in A1033 in the USSRH category. Together with the further steepening of the spectrum toward higher frequencies, this agrees with a cutoff in the electron spectrum as predicted by the turbulent acceleration model \citep{Brunetti2001,Petrosian2001,Kuo2003,Cassano2006,Brunetti2008}. 
It is well known that radio halos in more massive clusters are more powerful \citep{Cassano2013,vanWeeren2021,Cuciti2021}. Because the mass of A1033 is considerably lower than that of the vast majority of clusters that are known to host a radio halo, the cluster is expected to have a lower radio power.
In \autoref{fig:radiopower} we compare the 150\,MHz radio power of the A1033 halo to the radio halos and to the correlations of halo power to cluster mass presented in \citet{vanWeeren2021}. Furthermore, we show the two halos detected in the LOFAR deep fields \cite{Osinga2021} and the low-power halos discovered in \citet{Hoang2021} and \citet{Botteon2021}. All these radio halos are detected close to 150\,MHz and are hosted by clusters in the PSZ2 catalog. This comparison shows that while A1033 is one of the lowest-mass clusters hosting a radio halo, the halo is quite powerful. It lies considerably above the low-frequency correlations. Compared to the 150\,MHz correlations from \citet{vanWeeren2021}, it is overluminous by a factor in the range of 7 - 55. This is much larger than the intrinsic scatter of the correlations, which is $\sim3$.

The relatively large radio power is notable given the ultra-steep spectrum nature of the A1033 halo because in general, USSRH are expected to be less powerful \citep{Cassano2010b,Cuciti2021}.
We investigated whether the power estimate of the halo might be contaminated by other sources. We found that while the region we used to subtract the GReET indeed contains the significant emission in the HBA high-resolution map, images at lower resolution indicate a less confined feature with a ten times lower surface brightness in extension of the GReET. This patch of emission is coincident with the brightest region of the unobscured part of the radio halo, labeled \textit{P} in \autoref{fig:imghalo}. 
To ensure that patch \textit{P} does not contaminate our radio power estimate significantly in case it is not related to the halo, we repeated the flux density calculation, this time with a more conservative mask. The resulting radio power estimate is lower by only 1.3\%. Consequently, contamination from embedded sources cannot explain the large observed power.

A radio halo with a power above the correlation should correspond to an energetic merger event and/or a merger in an evolutionary phase close to the peak power. In both cases, the X-ray surface brightness distribution should show strong signs of disruption. The X-ray morphological parameters for the cluster were determined in the Chandra analysis in \citet{deGasperin2015}.
While the X-ray concentration parameter $c$ \citep[defined, e.g., in][]{Cassano2013} does not unambiguously qualify the cluster as a merging one ($c=0.200 \pm 0.004$), the centroid shift $w = 0.086 \pm 0.006$, the high differential velocity of the brightest cluster galaxy (BCG), and a distinct bimodality in the redshift distribution indicate a clear and possibly strong merger scenario \citep{deGasperin2015}. The strong merger might not fully show in the X-ray morphology parameters because it is likely occurring with a significant line-of-sight component, as also concluded in \citet{deGasperin2015} based on the BCG dynamics and redshift bimodality.
\citet{Cuciti2021} found a trend of the distance from the correlation of radio power to cluster mass and the X-ray morphological disturbance. Following their definition of the disturbance $d_\mathrm{X-ray}$, we have $d_\mathrm{X-ray}=0.86\pm0.03,$ which places A1033 in the region of disturbed clusters.
While a more energetic merger should lead to a flatter spectral index, the relatively low cluster mass might be the dominant factor determining the spectral properties. An increasing number of spectral studies of halos in low- and intermediate-mass clusters will help to constrain the spectral properties of halos in this mass regime.

A further reason for the comparatively high power and the steep spectrum of the halo in A1033 could lie in a greater abundance of seed electrons in the cluster, possibly accumulated by the various other bright radio sources. 
The synchrotron luminosity in reacceleration models is given by \citep[e.g.,][]{Brunetti2020,diGennaro2021}
\begin{equation} \label{eq:syncpower}
P\propto F \eta \frac{B^2}{B^2 + B_{\mathrm{cmb}}^2},
\end{equation} where $\eta$ is the (re-)acceleration efficiency, and $F=\rho v_{L_\mathrm{inj}}^3 / L_\mathrm{inj}$ is the turbulent energy flux in the emitting volume, which depends on the ICM density $\rho$ as well as the injection scale $L_\mathrm{inj}$ and turbulent velocity $v_{L_\mathrm{inj}}$. The efficiency $\eta$ is the fraction of the turbulent energy flux contributing to CRe acceleration and proportional to the CRe energy density $u_\mathrm{CR}$: $\eta \propto u_\mathrm{CR}$ \citep{Brunetti2007,Bonafede2022}.
Following from \autoref{eq:syncpower}, for the same CRe energy spectrum, a higher electron density would accordingly increase the halo power while not causing a spectral flattening. 
We speculate that at the low surface brightness patch in extension of the GReET (marked \textit{P} in \autoref{fig:imghalo}) with spectral properties in between those of the GReET and the halo, we might witness electrons that were conserved by the GReET entering the CRe population associated with the halo. A similar connection was proposed in \citet{Wilber2018}, for example. 

Because only a very small number of halos are detected in clusters with masses similar to or lower than A1033, the shape and scatter of the correlation in this mass range is not yet well determined. The detailed analysis of the largest sample of radio halos so far, based on the second data release of LoTSS \citep{Botteon2022}, will provide strongly improved constraints in the near future (Cuciti et al. in prep.).

\subsection{Radio phoenix}
\begin{figure}
    \centering
    \includegraphics[width=1.0\linewidth]{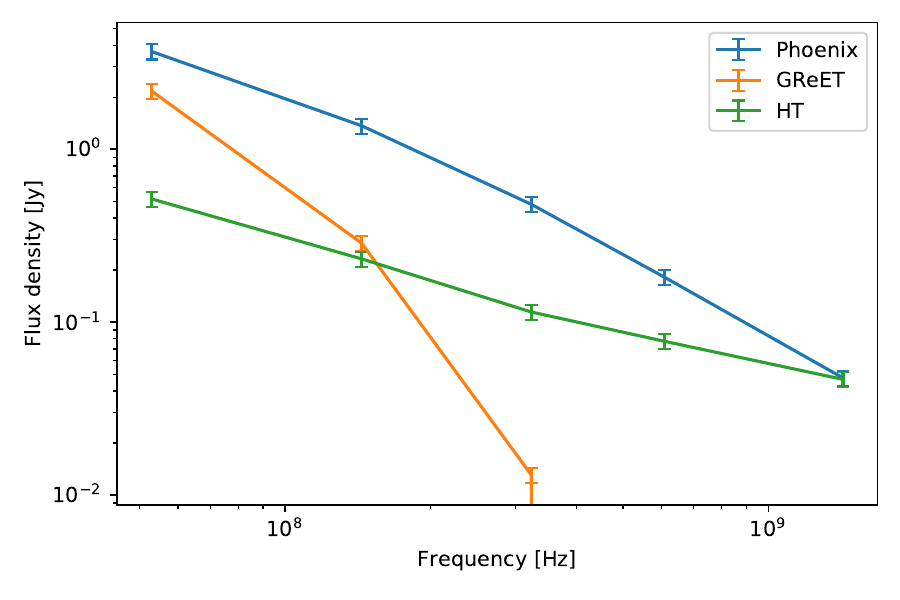}
    \caption{Flux densities as a function of frequency. Additional GMRT and VLA data at 609\,MHz and 1.4\,GHz are taken from \citet{deGasperin2015,deGasperin2017}.}
    \label{fig:measure}
\end{figure}

The bright extended source $150\,$kpc south of the cluster center was previously studied in \cite{deGasperin2015} and was categorized as a radio phoenix. A connection to a bright elliptical galaxy (marked \textit{S} in \autoref{fig:overview}) that coincides with a radio point source at 1.4\,GHz was proposed. The radio phoenix has an irregular and elongated shape with a largest linear size of $\sim 380$\,kpc. Toward the south, it shows a steep gradient, while toward the cluster center, the source is fading more slowly, and at lower frequencies, it extends toward the head-tail galaxy and the BCG. In \autoref{fig:measure}, its flux density is compared to the head-tail radio galaxy and the GReET between 54 and 1400\,MHz. The source shows spectral curvature. The spectral index flattens from $\alpha_{609}^{1425}=-1.58\pm0.17$ \citep{deGasperin2015} to $\alpha_{54}^{144}=-0.99\pm0.14$. The spatial variation of the spectral properties (\autoref{fig:si}) does not reveal a clear trend. The spectral curvature we observe is a characteristic feature of radio phoenixes \citep{Ensslin2001}. Therefore, the ultra-low frequency picture of the source is in line with the phoenix classification. The source differs from the GReET in the spectral shape, its spectrum is significantly less curved between 54 and 323\,MHz ($\Delta_\alpha\approx0.3$ to $\Delta_\alpha\approx2.0$), and furthermore, its morphology is more irregular and not clearly connected to a radio galaxy.    

\section{Conclusion}
We presented an analysis of the galaxy cluster Abell 1033 based on new ultra-low frequency data taken with LOFAR LBA at 54\,MHz and on existing data at higher frequencies (LOFAR HBA data at 144\,MHz and GMRT data at 323\,MHz). The cluster is an especially interesting target because despite its comparatively low mass, it hosts a variety of strong and extended radio sources such as a GReET, a radio phoenix, interacting AGN, and a radio halo, which show very heterogeneous properties.
Our findings are listed below.
\begin{itemize}
\item For the GReET, we detected extreme spectral curvature. The spectrum steepens form $\alpha\approx-4$ between 144 and 323\,MHz to $\alpha\approx-2$ between 54 and 144\,MHz. This finding is in line with the emission being generated by strongly aged electrons that are reaccelerated by a mechanism with a very low acceleration efficiency. 
\item Assuming a maximum lifetime scenario, we found a lower limit for the projected velocity of the WAT of 780\,km/s; this agrees with earlier findings. This corresponds to an maximum age of the tail of $\sim 800$\,Myr.
\item We investigated the properties of the radio halo in A1033. We found a radio power of $P_\mathrm{150\,MHz} \approx 1.22 \times 10^{25}$\,W\,Hz$^{-1}$. This power is higher than the correlations of radio power to cluster mass by a factor $>7$. At the same time, we found the spectrum of the halo to be ultra-steep between 54 and 144\,MHz ($\alpha \approx -1.69$), with a further steepening at higher frequencies. 
\item We presented possible reasons for both the high luminosity and the ultra-steep spectrum of the halo: The halo power might be explained by an energetic merger that does not fully reveal itself in the X-ray morphology parameters due to a significant line-of-sight component of the merger axis. This scenario is supported by the redshift bimodality and the high radial velocity of the BCG. This requires that the spectral properties are dominantly controlled by the cluster mass. Alternatively, a particularly high density of CRes, possibly accumulated by the various bright radio sources in the cluster, can explain an increased radio power without spectral flattening. Simulations could be used to investigate whether a significant per-cluster variation of seed electrons can be caused by sources such as the GReET and radio phoenix. 
\item We detected two candidate disrupted filaments in the GReET. We speculated that large-scale turbulent motions are responsible for the disruption. With the lower limit on the WAT velocity we derived, we constrained the turbulent velocity to be greater than $167\,\mathrm{km}\,\mathrm{s}^{-1}$ and $464\,\mathrm{km}\,\mathrm{s}^{-1}$, which is within the typical range found in simulations. The thin tail itself could be stabilized by magnetic fields if the turbulence at these small scales becomes sub-Alfv\'enic. 
\item We found that the spectrum of the radio phoenix flattens from $\alpha \approx -1.6$ to $\alpha \approx -1.0$ between 1.4\,GHz and 54\,MHz. This which is a characteristic property of radio phoenixes.
\end{itemize}
With the further advance of low-frequency radio surveys such as LoTSS and LoLSS, it will be possible to perform spectral analyses for a rapidly growing number of cluster radio sources. These will show whether the GReET in A1033 is representative for a greater category of sources or if it is truly special in terms of its low acceleration efficiency. Additionally, systematical analyses of radio halos in clusters of low and moderate mass will lead to a better understanding of the halo power and the spectral index distributions in this mass range.
\begin{acknowledgements}
We thank the referee for the useful comments which helped to improve the manuscript. LOFAR is the Low Frequency Array designed and constructed by ASTRON. It has observing, data processing, and data storage facilities in several countries, which are owned by various parties (each with their own funding sources), and which are collectively operated by the ILT foundation under a joint scientific policy. The ILT resources have benefited from the following recent major funding sources: CNRS-INSU, Observatoire de Paris and Université d’Orléans, France; BMBF, MIWF-NRW, MPG, Germany; Science Foundation Ireland (SFI), Department of Business, Enterprise and Innovation (DBEI), Ireland; NWO, The Netherlands; The Science and Technology Facilities Council, UK; Ministry of Science and Higher Education, Poland; The Istituto Nazionale di Astrofisica (INAF), Italy. This research made use of the Dutch national e-infrastructure with support of the SURF Cooperative (e-infra 180169) and NWO (grant 2019.056). The Jülich LOFAR Long Term Archive and the German LOFAR network are both coordinated and operated by the Jülich
Supercomputing Centre (JSC), and computing resources on the supercomputer JUWELS at JSC were provided by the Gauss Centre for Supercomputing e.V. (grant CHTB00) through the John von Neumann Institute for Computing (NIC). This research made use of the University of Hertfordshire high-performance
computing facility and the LOFAR-UK computing facility located at the University of Hertfordshire and supported by STFC [ST/P000096/1], and of the Italian LOFAR IT computing infrastructure supported and operated by INAF, and by the Physics Department of Turin university (under an agreement with Consorzio Interuniversitario per la Fisica Spaziale) at the C3S Supercomputing Centre, Italy. The data are published via the SURF Data Repository service which is supported by the EU funded DICE project (H2020-INFRAEOSC-2018-2020 under Grant Agreement no. 101017207).
This project is funded by the Deutsche Forschungsgemeinschaft (DFG, German Research Foundation) under project number 427771150. FdG acknowledges support from the Deutsche Forschungsgemeinschaft under Germany’s excellence strategy - EXEC2121 “Quantum Universe” - 390833306. V.C. acknowledges support from the Alexander von Humboldt Foundation. RJvW acknowledges support from the ERC Starting Grant ClusterWeb 804208. AB acknowledges support from the VIDI research program with project number 639.042.729, which is financed by the Netherlands Organisation for Scientific Research (NWO), and from the ERC-StG DRANOEL n. 714245. AD acknowledges support by the BMBF Verbundforschung under the grant 05A20STA. 

\end{acknowledgements}


\begin{appendix}
\onecolumn
\section{Uncertainty maps}
\FloatBarrier
\begin{figure}[ht!]
\centering
    \includegraphics[width=0.33\linewidth]{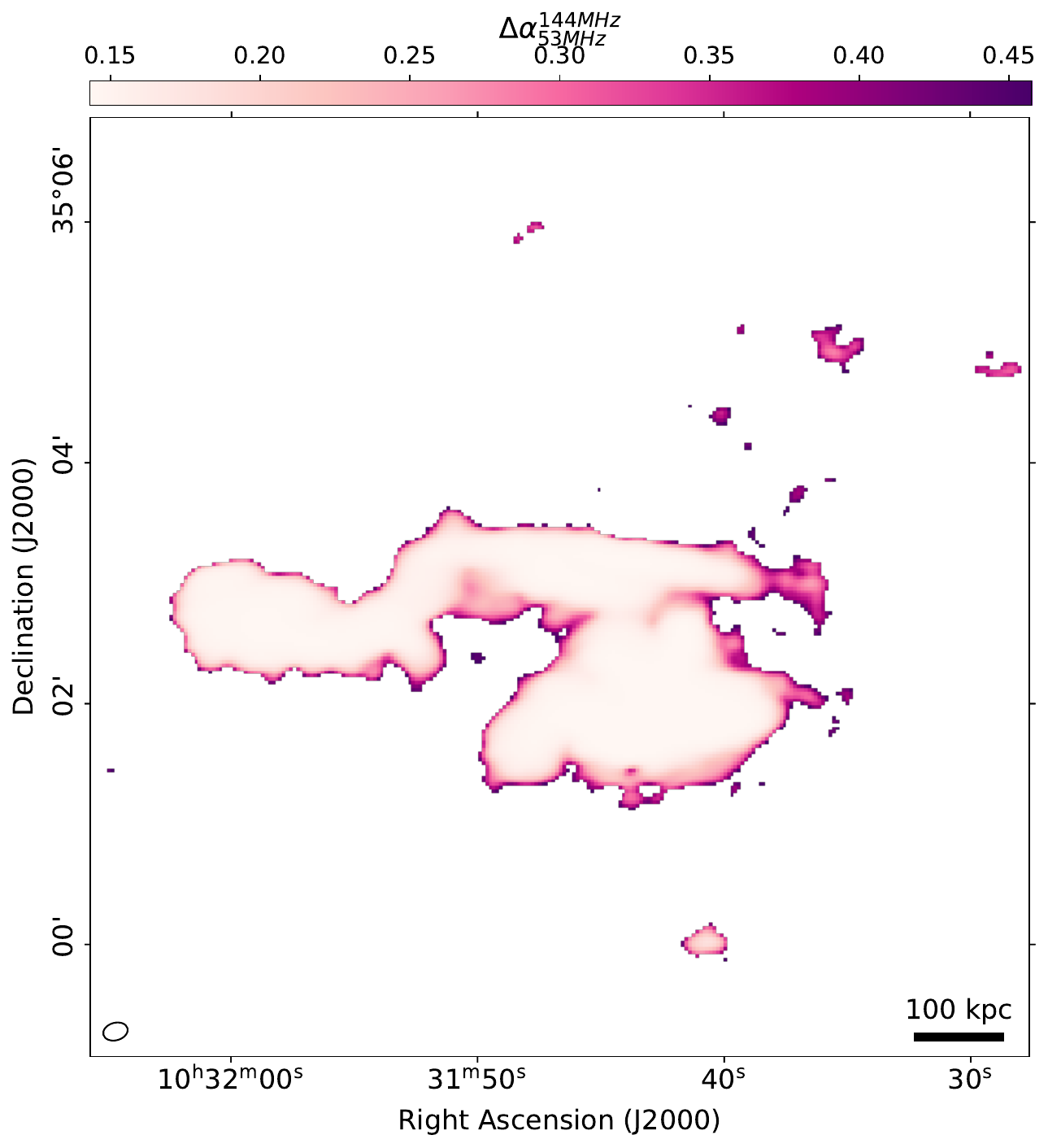}
    \includegraphics[width=0.33\linewidth]{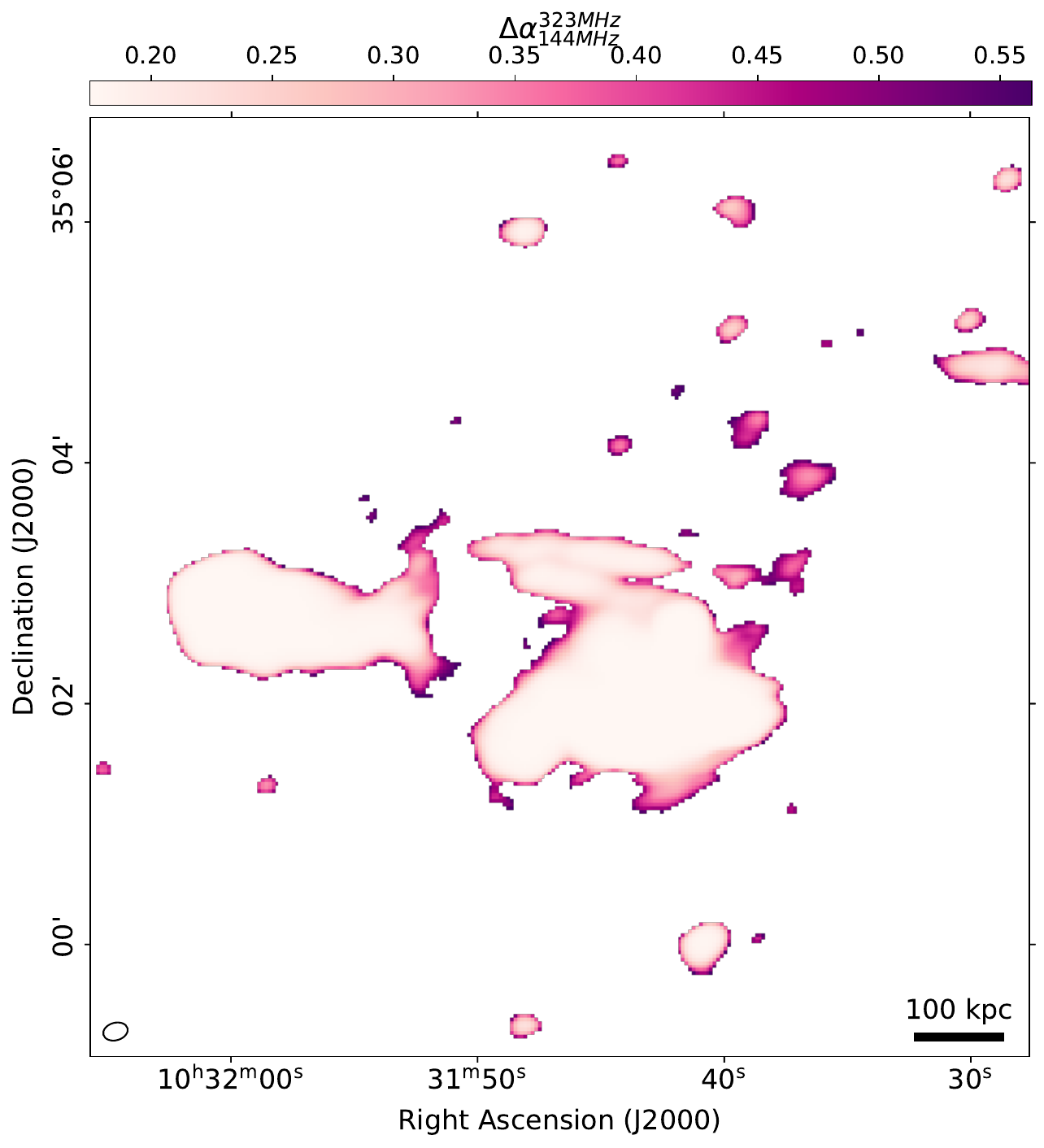}
    \includegraphics[width=0.33\linewidth]{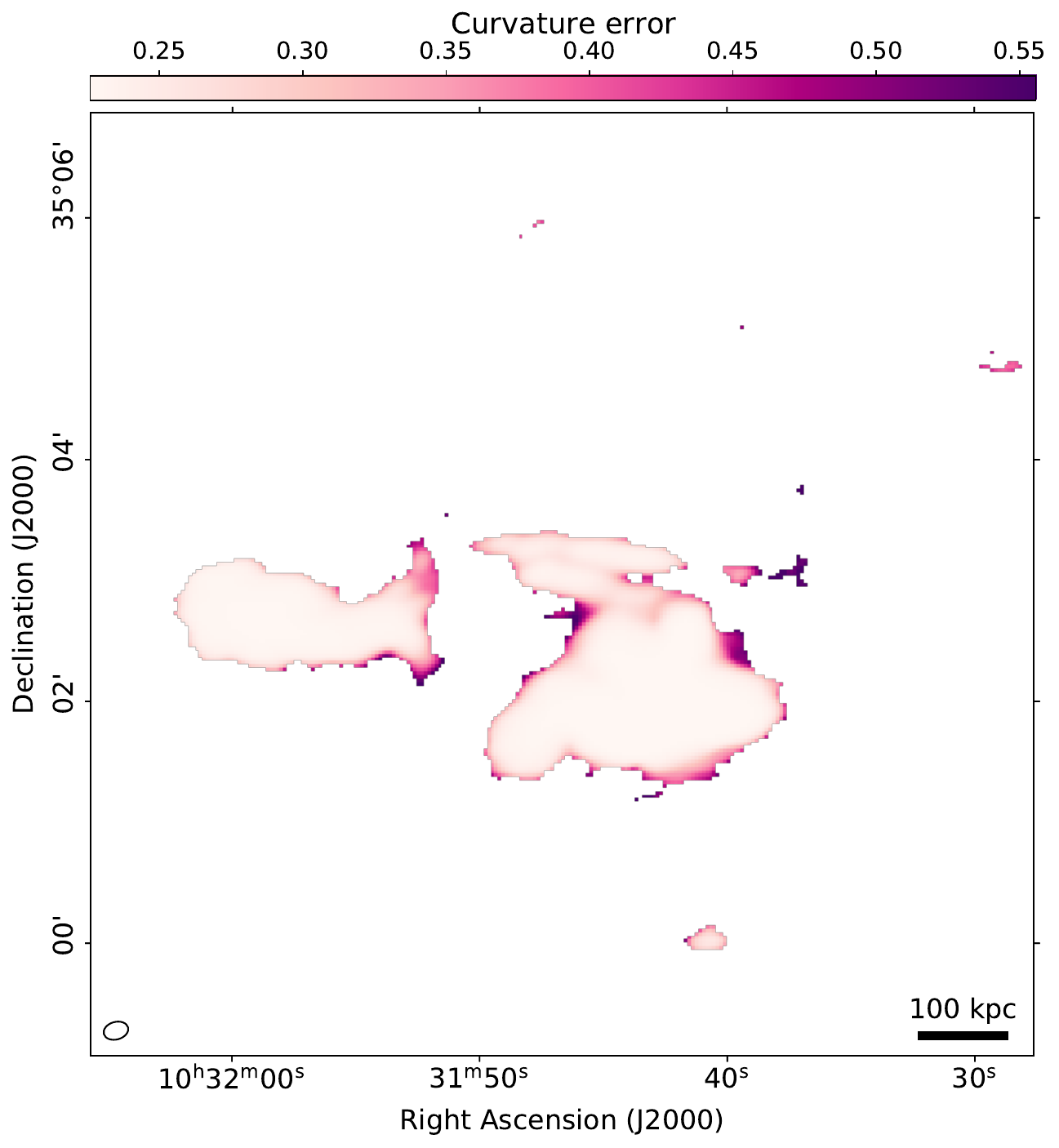}
\raggedright
\caption{Uncertainty maps to \autoref{fig:si}. The left and center figure show the spectral index uncertainties between 54-144\,MHz (left) and 144-323\,MHz (center). The right image shows the uncertainty of the curvature map.}\label{fig:sierr}
\end{figure}
\begin{figure*}[ht!]
    \centering
    \includegraphics[width=0.33\linewidth]{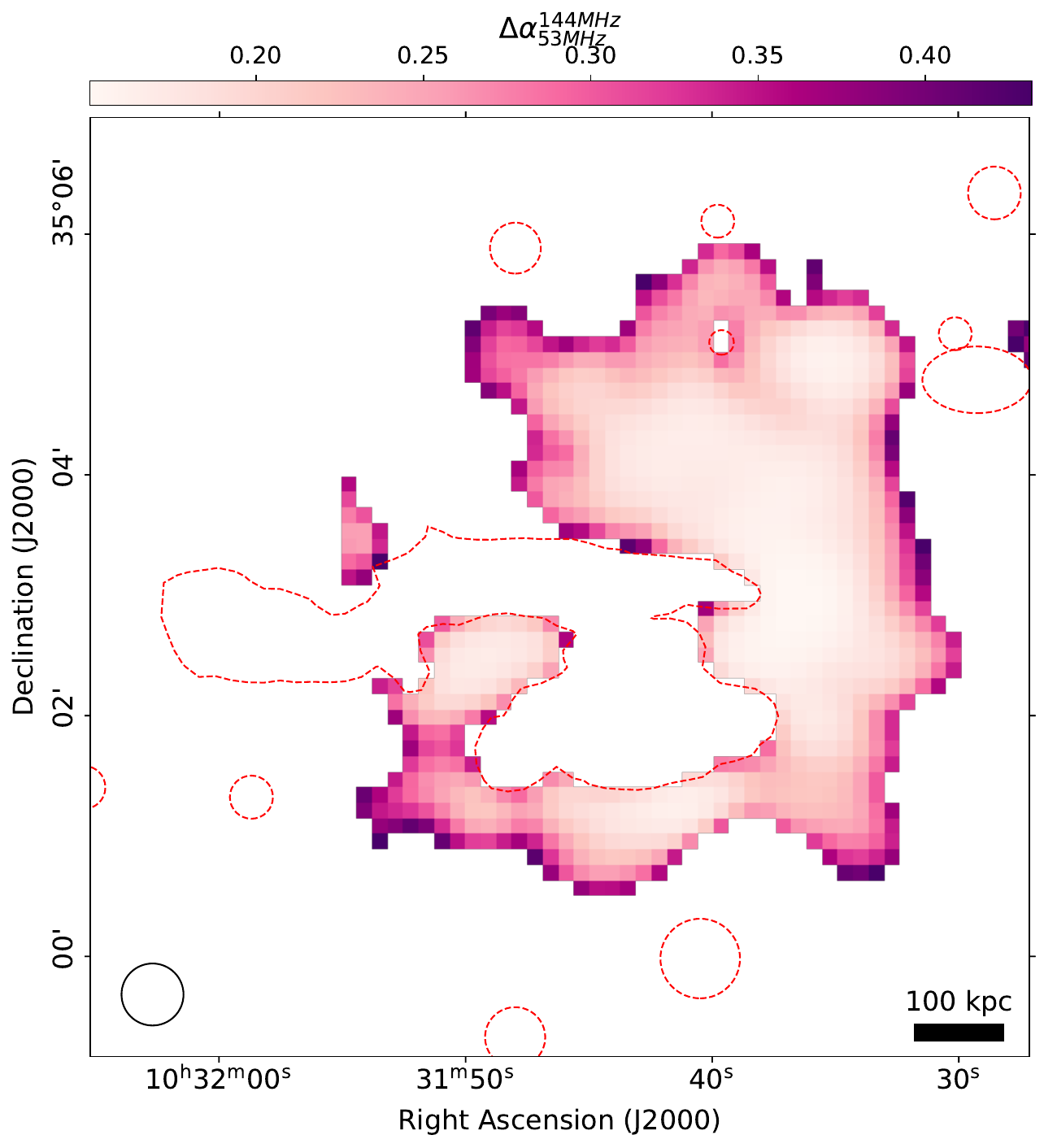}
    \includegraphics[width=0.33\linewidth]{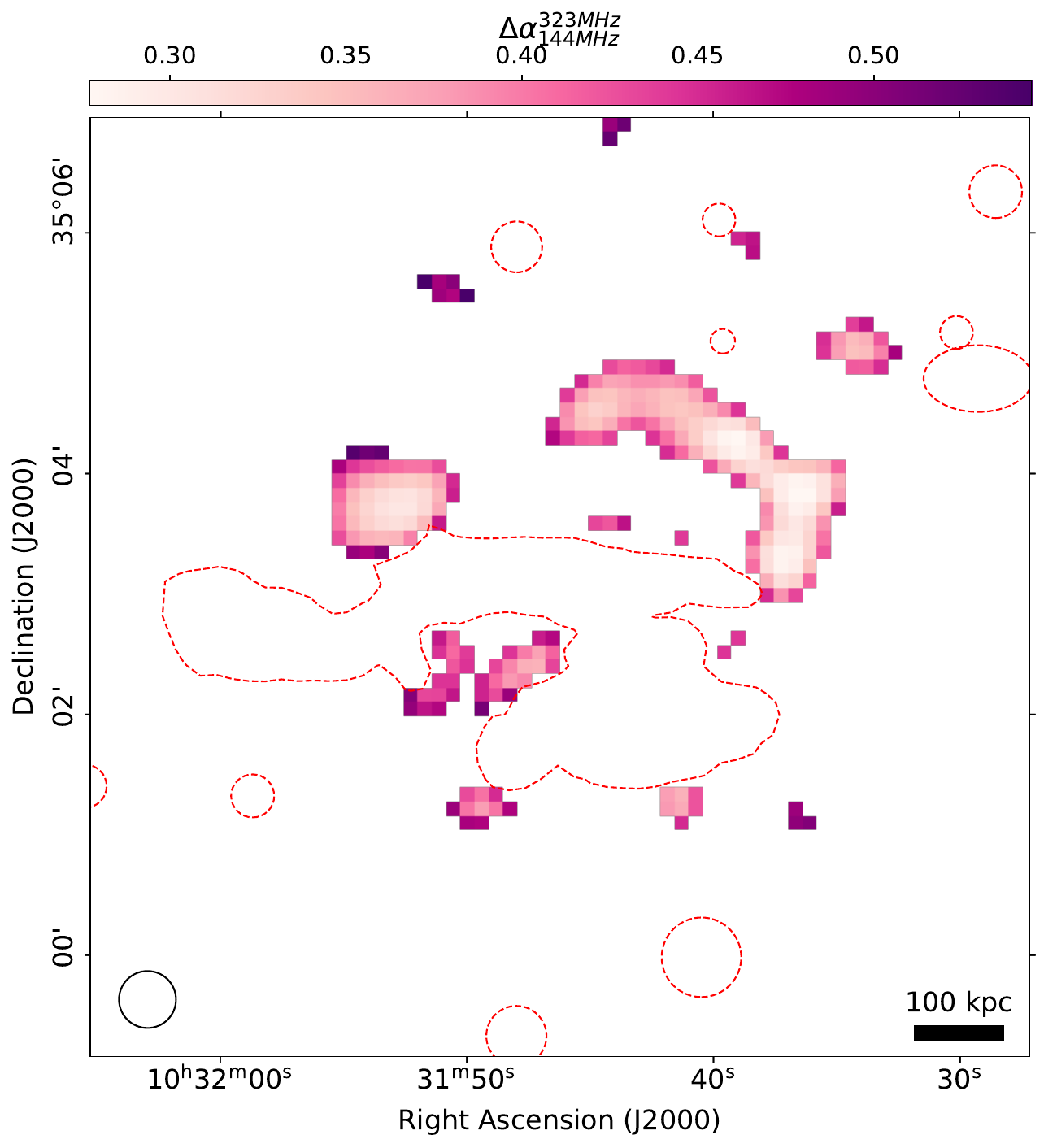}
    \caption{Source-subtracted spectral index uncertainty maps at a spatial resolution of 30''. The two figures are for the frequency ranges 54-144\,MHz (left) and 144-323\,MHz (right). Dashed red lines highlight the regions from which sources are subtracted. }\label{fig:sisubtracterr}
\end{figure*}
\FloatBarrier
\newpage
\section{Spectral aging model} \label{sec:model}
We employed the standard JP spectral aging model to describe the aging in the first section of the WAT radio galaxy. The flux density of this model at a frequency $\nu$ and after a time $t$ is given by \citep{Harwood2013}
\begin{equation}\label{eq:JPflux}
    S_\mathrm{JP}(N_0,\nu,B,t,z,\alpha_\mathrm{inj}) = N_0\frac{\sqrt{3} \ln(10) e^3 B}{16 \pi \epsilon_0 c m_\mathrm{e} (z+1)^2} \int_0^\pi \dif{\delta} \sin(\delta)^2 \int \dif{\log(E)} E F(x) n_\mathrm{e}(E,B,t,z,\alpha_\mathrm{inj}),  
\end{equation}
where $e$ and $m_\mathrm{e}$ are the electron charge and mass, $\epsilon_0$ is the vacuum permittivity, $\delta$ is the pitch angle, and $F(x)$ is the following function:
\begin{equation}
F(x) = x \int_x^\infty K_{5/3}(z)\dif{z}.    
\end{equation}
The variable $x=\nu/\nu_\mathrm{c}$ is the ratio of the frequency $\nu$ and the critical frequency $\nu_\mathrm{c}$ , which is given by
\begin{equation}
    \nu_\mathrm{c}= \frac{3 E^2 e B\sin(\delta)}{4\pi m^3_\mathrm{e} c^4}.
\end{equation} 
In \autoref{eq:JPflux}, the energy-integration is performed in log-space for numerical efficiency. We first fit a spectral index model to the observed spectral index values $\alpha$ to remove the dependence on the normalizations $N_0$. The spectral index model is calculated as 
\begin{equation}
    \alpha_{\mathrm{JP},i}({\nu_1,\nu_{2},B,t,z,\alpha_\mathrm{inj}}) = \frac{\log{\frac{S_\mathrm{JP}(\nu_1,B,t,z,\alpha_\mathrm{inj})}{S_\mathrm{JP}(\nu_2,B,t,z,\alpha_\mathrm{inj})}}}{\log{\frac{{\nu_1}}{{\nu_2}} }}.
\end{equation}
The uncertainty for the observed spectral index values $\sigma_\alpha$ is obtained from the uncertainties of the flux densities $\sigma_S$ according to Gaussian propagation of uncertainty,
\begin{equation}
    \sigma_\alpha = \frac{1}{\log{(\nu_1/\nu_2)}} \sqrt{\frac{\sigma_{S_1,\mathrm{stat.}}^2 + (0.1\times S_1)^2}{S_1^2}+\frac{\sigma_{S_2,\mathrm{stat.}}^2 + (0.1\times S_2)^2}{S_2^2}}.
\end{equation}
Here we take into account the 10\% systematic uncertainty on the flux scales.
We fit the WAT projected velocity $v_\bot$ by minimizing the standard $\chi^2$ statistics assuming a minimum aging magnetic field $B_\mathrm{min}$ and a linear motion $t = d_\bot/ v_\bot$,
\begin{equation}
    \chi^2 = \sum_{i=0}^9 \sum_{j=0}^1 \left(\frac{\alpha_i({\nu_j,\nu_{j+1}}) - {\alpha_{\mathrm{JP},i}({\nu_j,\nu_{j+1},B_\mathrm{min},d_\bot/ v_\bot,z,\alpha_\mathrm{inj}})}}{\sigma_{\alpha_i}({\nu_j,\nu_{j+1}})}\right)^2.
\end{equation}
Subsequently, a normalization factor $N_{0,i}$ is fitted in log-space for each beam-sized region along the WAT/GReET. 
\FloatBarrier
\begin{figure}[!ht]
    \centering
    \includegraphics[width=0.4\linewidth]{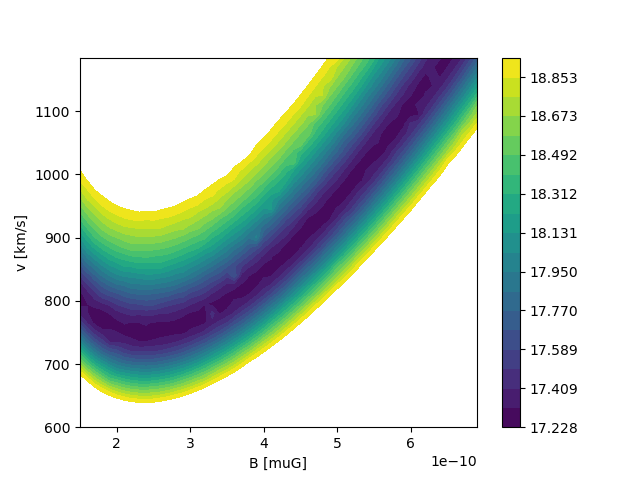}
    \caption{Aging model merit function against magnetic field ($x$-axis) and projected velocity ($y$-axis) in arbitrary units.}
    \label{fig:merit}
\end{figure}
\vfill
\FloatBarrier
\newpage
\section{Halo fit results}

\begin{figure*}[ht!]
\centering
\begin{subfigure}{1\textwidth}
    \centering
    \includegraphics[width=0.9\linewidth]{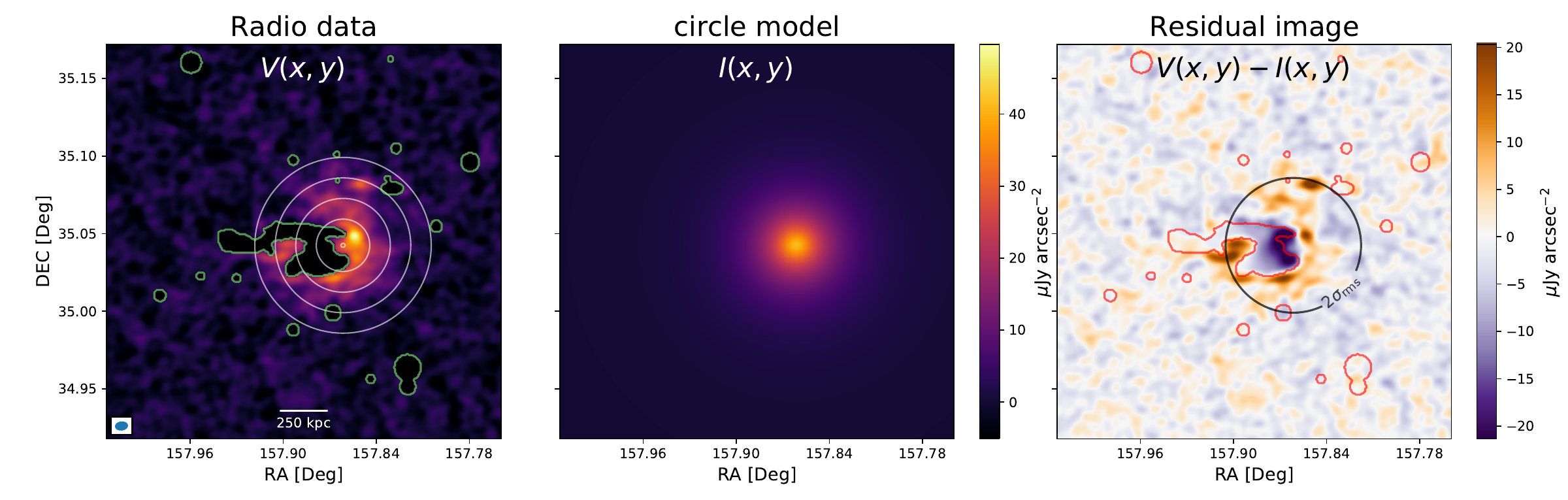}
    \caption{54\,MHz}
\end{subfigure}
\begin{subfigure}{1\textwidth}
    \centering
    \includegraphics[width=0.9\linewidth]{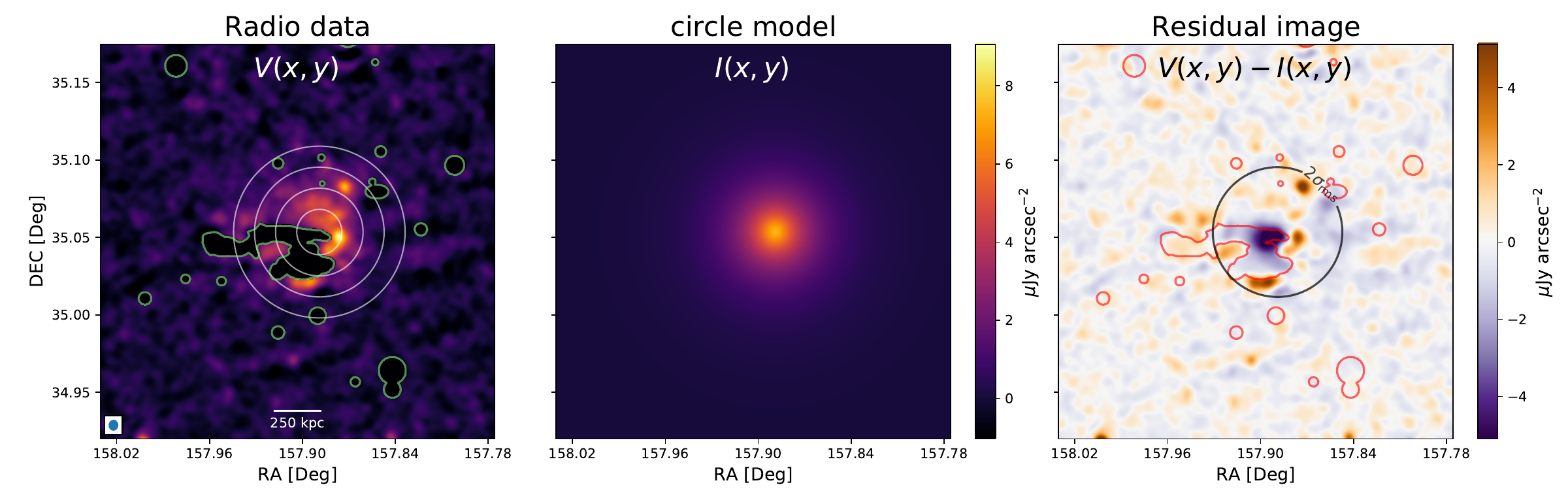}
    \caption{144\,MHZ}
\end{subfigure}
\caption{Results of the halo fitting in LBA (a) and HBA (b). The circles in the left panels indicate the position of the best-fit halo on top of the flux density maps. The center panels show the model, and the right panels present the residual. The green and red regions are excluded from the fitting. }\label{fig:halofit}
\end{figure*}%
\end{appendix}

\end{document}